\documentclass[manuscript]{aastex}
\usepackage{booktabs}
\usepackage{rotating}
\usepackage{fixltx2e}

\shorttitle{Local Bubble Survey}
\shortauthors{Farhang et al.}

\begin{document}
\title{Probing the Local Bubble with Diffuse Interstellar Bands. II. The DIB properties in the Northern hemisphere}

\author{Amin Farhang\altaffilmark{1,2}, Habib G. Khosroshahi\altaffilmark{1}, Atefeh Javadi\altaffilmark{1}, Jacco Th. van Loon\altaffilmark{3}, Mandy Bailey\altaffilmark{3}, Alireza Molaeinezhad\altaffilmark{1}, Saeed Tavasoli\altaffilmark{1}, Farhang Habibi\altaffilmark{1}, Ehsan Kourkchi\altaffilmark{1,5}, Sara Rezaei\altaffilmark{1}, Maryam Saberi\altaffilmark{1}, Liam Hardy\altaffilmark{4}}

\email{a.farhang@ipm.ir}

\altaffiltext{1}{School of Astronomy, Institute for Research in Fundamental Sciences (IPM), P. O. Box 19395-5746, Tehran, Iran}
\altaffiltext{2}{Department of Physics, Sharif University of Technology, P. O. Box 11365-9161, Tehran, Iran}
\altaffiltext{3}{Astrophysics Group, Lennard-Jones Laboratories, Keele University, Staffordshire ST5 5BG, UK}
\altaffiltext{4}{Isaac Newton Group, Apartado 321, E-38700 Santa Cruz de La Palma, Spain}
\altaffiltext{5}{Institute for Astronomy, University of Hawaii, 2680 Woodlawn Drive, HI 96822, USA}


\def \kms {\rm{km~s^{-1}}} 
\def \hii {\mbox{H\,{\sc ii~}}}
\def \hi {\mbox{H\,{\sc i~}}}
\def \na {\mbox{Na\,{\sc i~}}}
\def \ca {\mbox{Ca\,{\sc ii~}}}
\def \do {\mbox{$\lambda 5780~$}}
\def \dt {\mbox{$\lambda 5797~$}}
\def \bv {\mbox{E\,{\textsubscript {B-V}~}}}


\begin{abstract}
We present a new high signal to noise ratio spectroscopic survey of the Northern hemisphere to probe the Local Bubble and its surroundings using the \do \AA~and \dt \AA~Diffuse Interstellar Bands (DIBs). We observed 432 sightlines to a distance of 200 pc over a duration of 3 years. In this study, we establish the \do and \dt correlations with Na\,{\sc i}, \ca and E\,{\textsubscript{B-V}}, for both inside and outside the Local Bubble. The correlations show that among all neutral and ionized atoms, the correlation between Ca\,{\sc ii} and $\lambda5780$ is stronger than its correlation with $\lambda5797$, suggesting that $\lambda5780$ is more associated with regions where Ca$^{+}$ is more abundant. We study the \do correlation with $\lambda5797$, which shows a tight correlation within and outside the Local Bubble. In addition we investigate the DIB properties in UV irradiated and UV shielded regions. We find that, within and beyond the Local Bubble, \dt is located in denser parts of clouds, protected from UV irradiation, while \do is located in the low density regions of clouds.
\end{abstract}



\section{Introduction}
We are living inside a region within the Milky Way disk with extremely low neutral gas densities $n(\rm{H}) \sim 0.01 $ cm$^{-3}$ \citep{Bohlin75,Weaver77} and purportedly hot temperature ($T \sim 10^{6}$ K, for more discussion see \cite{Welsh09}). This high temperature was derived based on the distribution of diffuse soft X-ray background emission \citep{Snowden98}. This large cavity is known as the Local Bubble (LB) or the Local Cavity \citep{paresce84}. The evidence that led to the recognition of the Local Bubble comprises: (1) the lack of spectral hardening in the observed soft X-ray background that requires low values of neutral gas absorption \citep{Bowyer68}; (2) low values of interstellar extinction measured for stars within a distance of $\sim$ 100 pc, as compared with the extinction in sightlines of more distant regions in the interstellar medium (ISM) \citep{Fitzgerald68}; (3) observations of Ly-$\alpha$ absorption of nearby stars and \hi column densities, measured by \cite{Bohlin75} which showed that this region has a low density.

Most of the theoretical models of the Local Bubble formation favor the production of highly ionized atoms in collisional ionization equilibrium at a conductive interstellar cloud interface \citep{Slavin89}. High spectral resolution observations have revealed many small and partially ionized \ca gas clouds (local fluff) within 30 pc \citep{Crawford01}, thus we expect a significant number of such fluffs at larger distances along sightlines through the Local Bubble. Therefore the distribution of hot gas remains unresolved.

According to recent observations of neutral (\mbox{Na\,{\sc i}}) and ionized (\mbox{Ca\,{\sc ii}}) atoms, the Local Bubble is extended to a distance of $\sim$ 80 pc in the Galactic Plane and up to hundreds of pc into the Halo \citep{Vergely01, Lallement03, Welsh10}. Also 3D gas maps around the Local Bubble reveal that this cavity has a chimney like structure with a narrow opening towards the North Galactic pole (tilted at $l \sim 180^{\circ}$ and $b \sim +75^{\circ}$) and a wider opening in the Southern hemisphere \citep{Welsh10}. Since the Local Bubble and most surrounding cavities are supposedly filled with hot gas densities, such 3D distributions have been studied with diffuse X-ray background emissions for compare the morphology of the nearby cavities with the soft X-ray data \citep{Puspitarini14}.

The origin of the Local Bubble is still unknown. \cite{Crawford91} found that the majority of the absorptions toward stars in the Sco$ - $Cen association (170 pc) have negative radial velocities relative to the Local Standard of Rest (LSR). Therefore, most of the diffuse clouds in this direction are moving away from the Sco$ - $Cen association and are approaching us. Currently, the most plausible model is that the Local Bubble has been created by a number of successive supernovae in the nearby Sco$ - $Cen association. \cite{Smith01}, inspected one-dimensional multiple explosion models and found that this scenario requires 2$ - $3 supernovae to have occurred in the general vicinity of the Sun within the last few million years (probably 2$ - $5 million years ago).

We initiated a high signal to noise spectroscopic survey to map the Local Bubble through absorption in Diffuse Interstellar Bands. ISM maps of DIBs have been created in the recent past, towards the globular clusters $\omega$ Centauri \citep{Jacco09} and the Tarantula Nebula in the Large Magellanic cloud \citep{vanLoon13}. Also the pseudo three$-$dimensional map of the 8620~{\AA} DIB within 3 kpc from the Sun have been produced \citep{Kos14}. In the first paper in this series (\cite{Bailey14} in preparation) we present the results from the Southern hemisphere survey. In this paper, we present the results from the Northern hemisphere survey. Our aim is to (I) investigate the DIB distribution within and outside the Local Bubble, and (II) examine the DIB correlations with atomic and ionized species in this region. We begin with a brief discussion in section~\ref{dib} of the history and the properties of DIBs. In section~\ref{observ} we present the technical aspects of the study, observations and the target selection. In section~\ref{dataanal}, we present the data analysis procedures. In the next three sections, we discuss the results of various DIB properties inside and outside the Local Bubble, according to their Equivalent Widths (EWs), correlations and UV sensitivity. Finally, in section~\ref{discus}, we summarize the implications of our results.

\section{Diffuse Interstellar Bands}
\label{dib}
Observing the first Diffuse Interstellar Bands, in the spectra of reddened stars, by \cite{Heger22} was the starting point to the oldest puzzle in stellar spectroscopy that still remains unresolved. While Heger argued that DIBs are stellar, more than a decade later \cite{Merrill36} argued for the interstellar nature of DIBs. At the present, the DIBs are known to comprise $\sim 500$ narrow to broad interstellar absorption features which are observed ubiquitously between 4000 and 10000 \AA~\citep{Herbig95}.

DIBs are studied toward specific targets or large numbers of stars. \cite{Hobbs09,Hobbs08} carried out a complete survey of HD 204827 and HD 183143 to investigate very weak and relatively narrow bands with typical EWs down to a few m\AA. Over the last decade many astronomical and laboratory experiments have been accomplished to find and understand DIB carriers as well as a number of theoretical studies. Based on these studies, DIB candidates could be among an infinite number of large carbon-based  "organic" molecules \citep{Sarre06}. To extract physical information about DIB carriers, some correlation studies have been done: (1) DIB correlations with atomic and molecular interstellar species, for instance many studies have considered the \do \AA~and \dt \AA~DIBs correlation with \hi, H$_{2}$, \na D lines, {Ca\,{\sc i}} (4227.9 \AA), \ca (3934.8 \AA), {K\,{\sc i}}, CH (4300.3 \AA), CH$^{+}$ (4232.3 \AA), CN (3874.6 \AA) and {Ti\,{\sc ii}} \citep{Herbig95,Weselak08}, (2) DIB correlations with extinction \bv \citep{Vos11} and (3) DIB correlations with other DIB absorptions \citep[see][]{Vos11,Friedman11,Kos13}.

DIBs are used for different purposes in astronomy. For instance, the most outstanding DIBs in the optical waveband; $ \lambda $4429, $\lambda$5780 and $\lambda$5797 are empirically known to trace the neutral phase of the ISM. The strength of these DIBs are correlated with the reddening E\,{\textsubscript {B-V}}, the neutral hydrogen and the \na column densities \citep{Herbig93,Friedman11}. $\lambda5780$ is well-correlated with column density of neutral hydrogen N(H); also, for Galactic sightlines which are not in high radiation environments, the DIB correlation with N(H) is stronger than the correlation with E\,{\textsubscript {B-V}}. Therefore  it is possible to estimate N(H) based on measurement of the equivalent width of \do \citep{Friedman11}.

There are still ambiguities about DIB carriers, for instance the large widths of DIBs argue that the carriers could not be atoms in the gas phase. On the other hand, by observing DIB features with high resolution spectrographs, some substructures appear in DIB profiles which indicate the carriers are large gas-phase molecules \citep{Sarre95}. In addition, DIBs react to the strength of the UV field of the local environment \citep{Cox06} and DIB strength varies as a function of UV radiation toward different sightlines \citep{Vos11}. For extra$-$galactic sightlines, observations show that DIBs are present only on sightlines with a significant UV bump \citep{Cox06}. Therefore, specific groups of UV resistant molecules, such as Polycyclic Aromatic Hydrocarbons (PAHs), fullerenes and carbon chains are commonly the more acceptable candidates for DIB carriers \citep{Herbig95}.

To further our understanding of the DIB carriers and study the physical properties of interstellar environments it is important to study DIBs behavior in different ISM regions. DIB carriers have been ubiquitously detected everywhere, for instance, DIBs have been observed in the Magellanic Clouds \citep{vanLoon13}, M31 and M33 \citep{Cordiner08a,Cordiner08b} and beyond the Milky Way in SN host galaxies \citep{Cox08}. Thearefore we study the DIB carriers within the purported hot Local Bubble and beyound it for investigating the DIB behaviours in such environment. 

\section{OBSERVATIONS}
\label{observ}
\subsection{Spectroscopy with IDS}
Observations were carried out over 35 nights on 10$-$15 October 2011, 1$-$6 March 2012, 8$-$12 September 2012, 31 December 2012 to 9 January 2013 and 27 May to 2 June 2013. All observations were carried out using the Intermediate Dispersion Spectrograph (IDS) at the 2.5 m Isaac Newton Telescope (INT) at the Roque de Los Muchachos in La Palma. The IDS employs a long-slit spectrograph with a set of 16 gratings and two CCDs: The blue-sensitive EEV10 CCD, and the RED+2 detector which is more sensitive in the red; both CCDs have $4096\times2048$ pixels. One could achieve dispersions between 0.24 and 4 \AA$/$pix. The spatial scale for the EEV10 is $ 0''.4$ and for the RED+2 is $ 0''.44 $ $/$pix, and the full unvignetted slit length is $ 3'.3 $. 

For our observations we used the 235 mm camera and H1800V IDS grating for an effective resolution of 0.31 \AA$/$pix. H1800V was chosen since it provides a high spectral resolution well matched to the typical width of DIBs. We chose 5800 \AA~as the central wavelength, in order to cover the major $\lambda$5780, $\lambda$5797, $\lambda$5850, $\lambda$6196 \& $\lambda$6203 DIBs. A $ 1''.1 $ slit yielded spectra in the 5750$-$6040 \AA~region at spectral resolutions of $R\equiv \lambda / \Delta \lambda \sim2000$ (or a velocity resolution of $\Delta v = 150$ km s$^{-1}$). The DIB detection requires a high signal-to-noise (S/N) ratio of at least 100 but for detection of very weak absorptions, like the one which was detected by \cite{Cordiner06} towards the nearby star $\mu^{1}$Cru with \do DIB equivalent width of 4 m\AA, we need S/N of at least 2000. The seeing varied during these observing nights from 1\arcsec.1 to 1\arcsec.9.

Each observing night we obtained more than 60 flat field frames (quartz lamp) with exposure times of 13$-$15 sec to correct the CCD pixel-to-pixel sensitivity variations and eliminate spurious pixels in the frames. For wavelength calibration, more than 15 arc frames (CuAr+CuNe) were taken at the beginning and the end of each observing night. A large number of bias frames were also recorded. We observed the targets with 9$ - $25 science frames for an individual star with an exposure time between 3 to 400 sec depending on the apparent magnitude of the stars to achieve a typical continuum S/N $ \sim$2000.

Some additional steps were taken in IRAF prior to extracting the spectrum with the KPNOSLIT package. The data were processed using the CCDRED data reduction package. First, an average bias frame was produced using the {\sc imcombine} task by stacking $\sim 100$ bias frames. Then the resultant frame was subtracted to correct a number of poor but recoverable columns.  To correct the CCD sensitivity pattern, we combined flat frames and produced a normalized master flat and divide each science frame by this master flat. After extracting the spectrum and calibrating the wavelength range with arc frames,  to increase the S/N ratio up to $\sim 2000$, by using {\sc scombine} task we combined the science frames. The overall S/N ratio in our study is $\sim 2000$ but in some cases we observed more than 25 frames, and by smoothing the spectrum by a $3\times3$ pixels box the S/N reaches to $\sim 2800$.

\subsection{Target selection}
\label{targets}
\begin{figure}
	\includegraphics[scale=0.45]{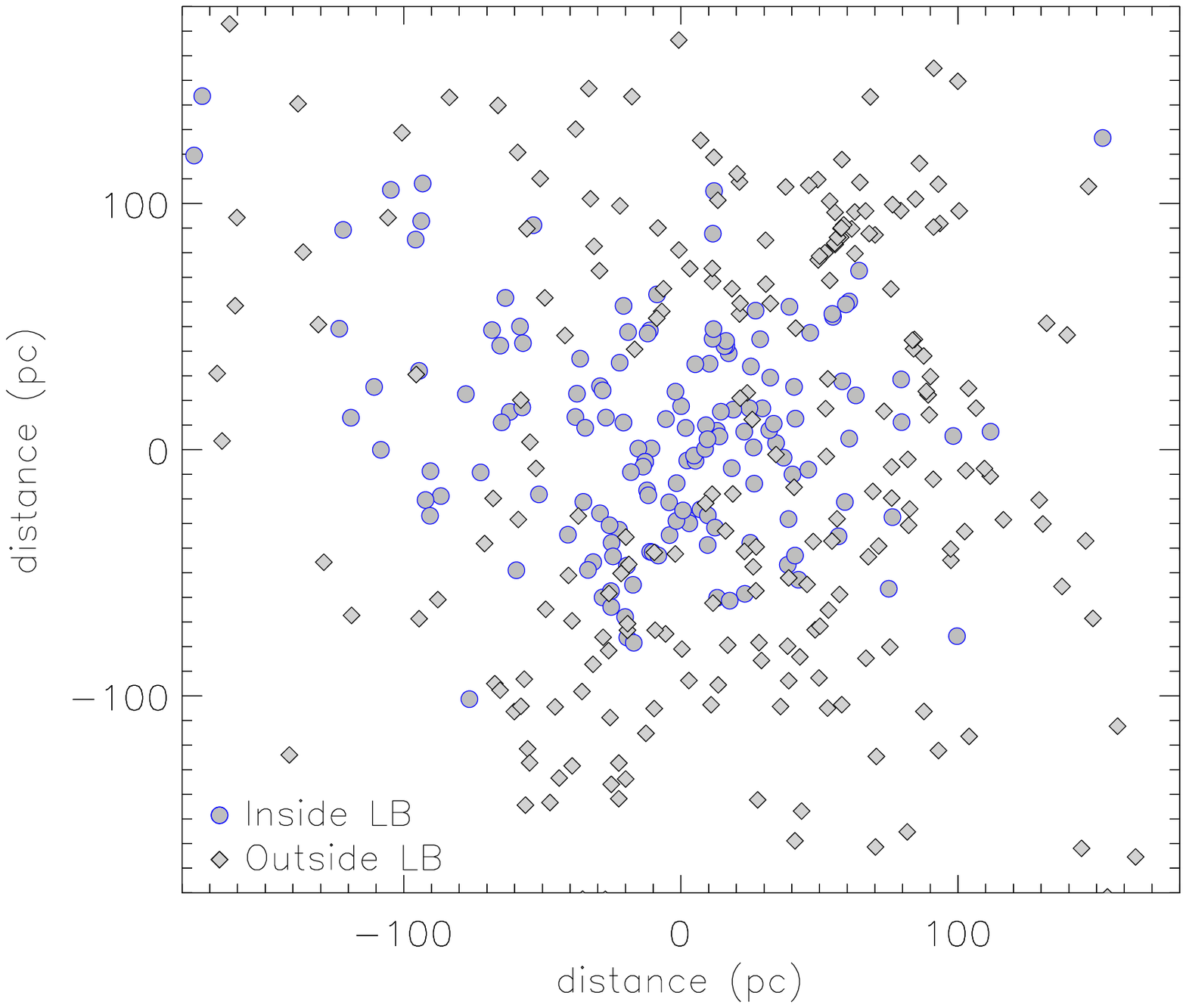}
	\caption{The projection of spatial distribution of observed stars around the Sun (map center) on the celestial equator plane. The circles are objects located inside the Local Bubble, and the diamonds show the observed targets situated outside the Local Bubble (see section \ref{targets} for details).}
	\label{fig1}
\end{figure}

\begin{figure}
	\includegraphics[scale=0.36]{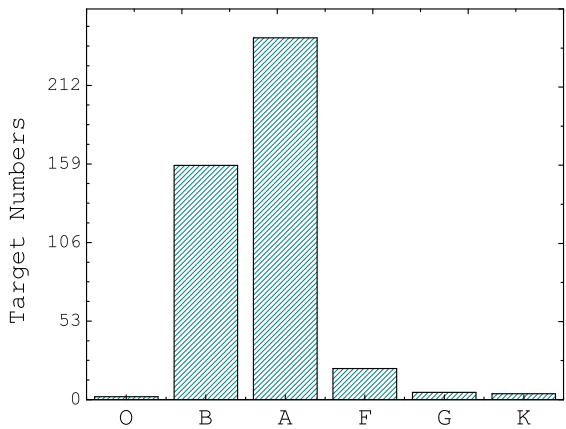}
	\caption{Histogram of observed stars spectral types. The B and A types represent more than 93\% of the observed stars, but for more sky coverage some cooler stars were observed.}
	\label{fig2}
\end{figure}

All 432 targets have been selected from the 3D Na I D lines survey of \cite{Welsh10}. The detailed list of targets are presented in Table 1 of the third paper in this series \citep{Farhang14}. In addition, all of the selected objects have well-known distances from the Hipparcos satellite \citep{Perryman97}. Since we wanted to observe the Local Bubble and its surroundings, we have selected bright stars up to a distance of 200 pc (Fig.~\ref{fig1}). To maximize the uniformity of the observed stars in the map and increase its covered density area, as well as to observe hot stars (O, B types) we observed some cooler stars (A, F, G \& K). As we show in Fig.~\ref{fig2}, the majority of observed stars are B and A types. 

\begin{figure}
	\includegraphics[scale=0.43]{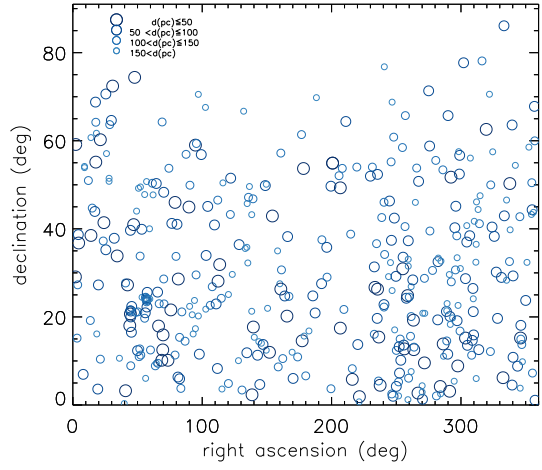}
	\caption{Observed stars in celestial coordinates. For covering all sky in RA the observations were done in 6 months intervals in three years.}
	\label{fig3}
\end{figure}

Since the saturation level of the IDS detector was $\sim 64000$ counts, a small number of targets were rejected due to their extreme brightness ($V < 1.8$ mag). To increase the observing run efficiency, targets with $V > 7.2$ mag were also rejected from our target list to avoid very long integration times. The pointing range of the INT (Zenith distance $< 70^{\circ}$ and declination range $> -30^{\circ}$), enabled us to observe some Southern celestial hemisphere targets, however our targets were predominantly selected from the Northern hemisphere. As shown in Fig.~\ref{fig3} our observations cover all of the Northern hemisphere. A few targets were acquired to overlap with the Southern sky survey to check the consistency of the measurements.

In this paper we perform our analysis separately for sightlines inside and outside the Local Bubble. But given the 3D shape of the Local Bubble, a chimney like structure, we could not divide inside and outside only based on the distance from the Sun (for example 80 pc in every direction). For selecting the stars located inside the Local Bubble firstly we select the objects with EW(\na D$_{2}$) $\leq 5$ m\AA, because within the Local Bubble the \na is very weak \citep{Welsh10}. Then according to their 3D map of \na and Ca\,{\sc ii}, we visually identify which one is located inside the Local Bubble chimney.

\section{Data analysis}
\label{dataanal}
\subsection{Fitting procedure}
\label{ewmeas}

Quantifying the equivalent widths for DIB absorption features is a real challenge in DIBs studies. In some cases, the DIB feature is blended by that of other chemical species. Since the DIB carriers are unidentified, the shape and the width of the spectral profiles remain uncertain. Besides overlap with other species, there is a possibility for blending with features from higher rotational levels of the same species \citep{Friedman11}. The equivalent width is defined as:

\begin{equation}
W = \int \frac{I_{0}(\lambda)-I(\lambda)}{I_{0}(\lambda)}~d\lambda = \int (1-\exp(-\tau(\lambda)))~d\lambda
\label{eqew}
\end{equation}

In this equation, the $I_{0}$ and $I_{\lambda}$ are fluxes of the continuum and the spectral line respectively. Also this definition shows that the EW of the line is proportional to the optical depth of the observed gas cloud $\tau(\lambda)$. For producing a normalized spectrum, continuum fitting to the observed spectra was performed by fitting a nine-order Legendre polynomial.

Since the DIB profile is inferred from an overlapping unknown number of atomic absorptions, the overall shape of the DIB profile is unknown. However the equivalent widths, line widths and the central velocities of the \do and \dt lines can be approximated by a Gaussian function \citep[e.g.,][]{Jacco09}. Therefore we obtain the line width in terms of the $\sigma$ value of the Gaussian distribution, and accordingly the Full-Width at Half Maximum (FWHM) is calculated as FWHM=$2(2~\ln2)^{1/2} \sigma = 2.355 \sigma$.

\subsection{Error estimation}
The principal known source of uncertainties on DIB's EW are the blending with the stellar lines and other DIBs. One of the sources of blending contamination for \do is the $\lambda5778$ DIB feature \citep{Herbig75} and the probable source of blending to \dt is the $\lambda5795$ DIB \citep{Krelowski97}. In the \do case, because of the difference of FWHM for these two DIBs the separation is easy but for \dt it is difficult \citep{Galazutdinov04}. The corresponding statistical uncertainty for each observed DIB is computed by standard deviation of residuals of the Gaussian fitting and summed in quadrature and weighted by the Gaussian fit \citep{Jacco09,Vos11}. The statistical error is always underestimated in DIB studies, since the main source of error in the equivalent widths is the error in finding the real position of the continuum (systematic error). For computing this error, we fit three different continuum lines to the local absorption region with $\pm 12$ \AA~range around the central DIB wavelengths (linear fit to the continuum, quadratic fit to the continuum and fit to DIB with linear continuum \citep{Kos13}). Accordingly, we set the intersection points of the DIB absorption and the continuum level \citep{Krelowski93}, and then compute equivalent widths. Therefore, the error for systematic uncertainty is determined by the difference between the highest and lowest values of equivalent width among these three EWs. 

\begin{figure*}
	\includegraphics[scale=0.8]{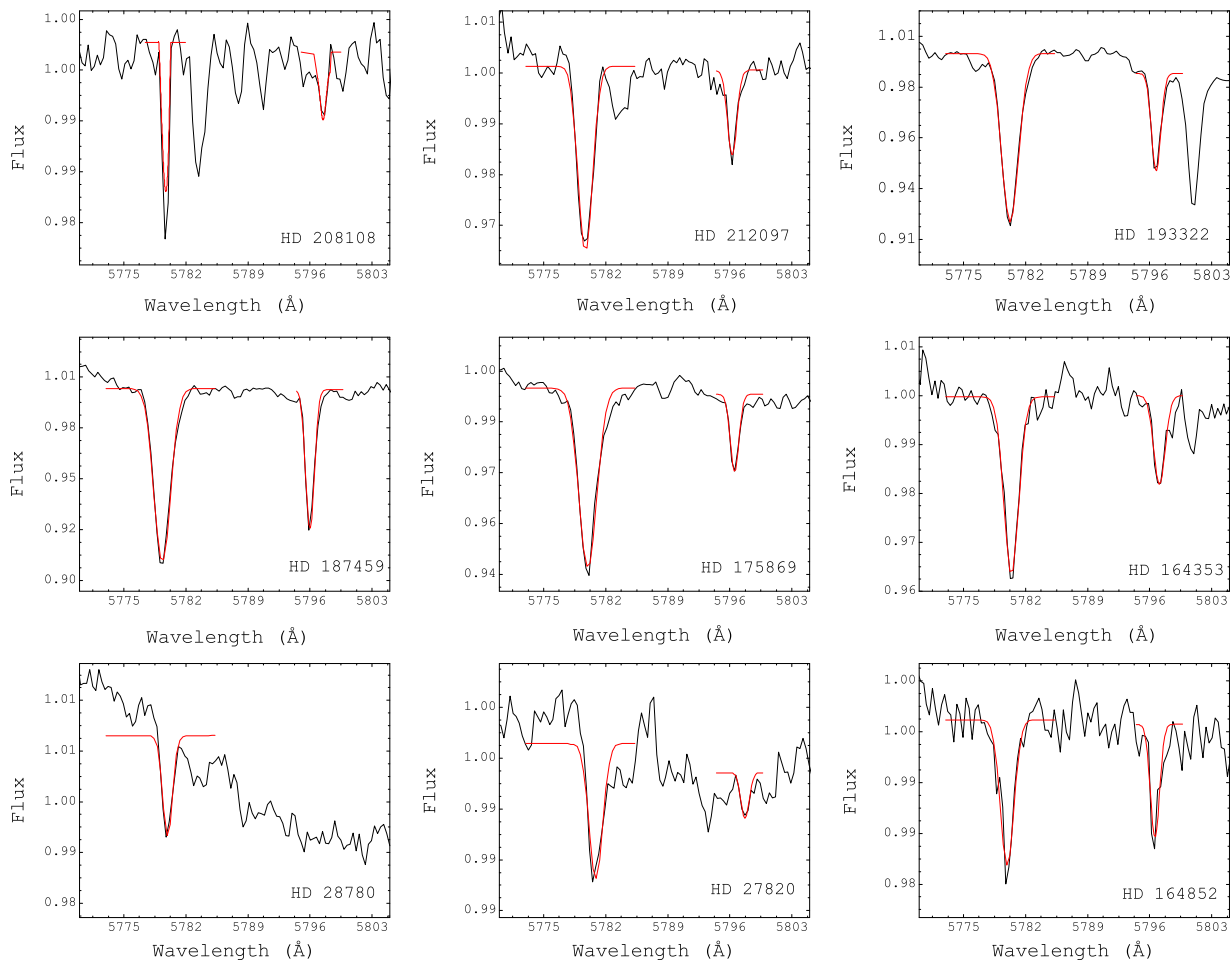}
	\caption{Some observed DIB features for quality checking. The first panel in the most upper left side (HD 208108) is an example with stellar absorptions, for which after subtracting a synthetic spectrum the Gaussian fit to \do was done. The lower left panel (HD 28780) is a sightline with observed \do and absence of \dt DIBs, which is common in DIB observations. The lower right panel (HD 164825) is an example with poor seeing that caused more noise than normal.}
	\label{spc}
\end{figure*}

As an example of the quality of the observed data, in Fig.~\ref{spc} we show some typical \do and \dt absorptions with the best Gaussian fit line to each one. In our observations the equivalent width of \do varies between 3 to 250 m\AA~but for some sightlines (e.g., HD 183143) the EW is near 800 m\AA. In our dataset the \dt EW varies between 3 to 150 m\AA.

\subsection{Synthetic stellar models} 
Extracting interstellar spectrum features using the observations of cool star spectra containing stellar absorption lines, requires a very accurate synthetic spectrum of the target star to be subtracted from the observed spectrum.

To model a stellar atmosphere, the input parameters are effective temperature $T_{{\rm eff}}$, surface gravity $\log~g$ and micro-turbulence velocity $\xi$, which leads to calculating the spatial distribution of some physical quantities like: temperature $T(r)$, electron density $n_{e}(r)$, population numbers $n_{i}(r)$, density $\rho(r)$, velocity field $v(r)$ etc.

We calculate our atmosphere models with the ATLAS9 code, with correction of iron and iron-peak element opacities \citep{Kurucz92}. We used ATLAS9 for modeling the stellar atmosphere in our samples, from \cite{Castelli03}\footnote{http://wwwuser.oats.inaf.it/castelli/} who use new Opacity Distribution Functions (ODFs) for several metallicities. Also increase spatial resolution to 72 plane parallel layers from $\log\tau_{{\rm Ross}} = -6.875$ to $+2$ and update all models with solar abundances \citep{Castelli97}. For modeling each observed target atmosphere we need the effective temperature of the observed star and the surface gravity $\log g$. We obtained these values from \cite{Cayrel96, Varenne99, Reddy03, Prugniel07,Huang10,Prugniel11} and \cite{Koleva12}. Also we determine target metallicity from \cite{Heacox79, Abt02, Royer02, Royer07} and \cite{Schroder09}. After modeling the stellar atmosphere we generate the synthetic stellar spectrum with the SYNTHE suite codes \citep{Kurucz05}, but we used the Linux port \citep{Sbordone04, Sbordone05}. Atomic and molecular data were taken from the data base on Kurucz's website\footnote{http://kurucz.harvard.edu} \citep{Kurucz05}. Also for considering the broadening caused by rotational velocity, we use the rotational velocity from \cite{Abt02, Royer02, Royer07} and \cite{Schroder09}, and for those stars for which the rotational velocity was not reported we use $v \sin i=15-20$ km s$^{-1}$. In our calculations we included all the atomic and molecular lines with empirically determined atomic constants plus all the diatomic molecular lines (CH, NH, CN, MgH, SiH, SiO, H$ _{2} $, C$_{2}$ and CO) except the TiO molecule.

\subsection{Mid-hot stars}
For all targets we carefully searched for the potential contaminating stellar lines. \cite{Friedman11} showed that for late-B objects there are three \mbox{Fe\,{\sc ii}} stellar lines (at 5784.45 \AA) that would contaminate the \do DIB, but this can be easily detectable. Therefore for our late-B type targets we correct the \do DIB absorption according to the above method. On the other hand, the main sources of A-type star contaminations at \do are two strong \mbox{Fe\,{\sc ii}} absorptions at 5780.13 \AA~and 5780.37 \AA~and a nearby \mbox{Fe\,{\sc ii}} at 5783.63 \AA. There is no source of contamination at $\lambda5797$. 

\begin{figure}
	\includegraphics[scale=0.38]{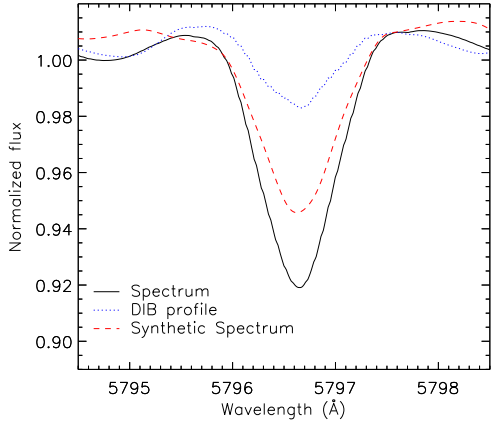}
	\caption{The black line is the observed spectrum including both stellar and interstellar absorptions, the dashed line is a synthetic stellar spectrum and the dotted line is the interstellar spectrum obtained by subtracting the synthetic spectrum from the observed one.}
	\label{dibsub}
\end{figure}

For some observed A-type stars that the metallicity have not been reported, we produce all possible synthetic spectra to compare with the observed spectra. The surface gravity of A-type stars varies between 3.5 to 4.2 \citep{Gray92}, therefore we choose a constant $\log g = 4$ in all the atmospheric models. From A0 to A9 spectral types, according to \cite{Theodossiou91} report, we select a constant temperature for luminosity classes I, II$-$III and IV$-$V. Also the A-type stars have different metallicities from 0 to $ -2 $ \citep{Beers01}. Accordingly, for a given A-type subdivision (e.g., A0) and luminosity class (e.g., IV), we produce three different atmosphere models with [Fe/H] $= 0, -0.5, -1.5$, and compare with the observed spectrum to choose the best model. Also we consider the effect of rotational velocity convolved with the instrument dispersion. But the rotational velocity of our observed A-type stars are very high ($\sim 200$ km s$^{-1}$) \citep{Hoffleit95}, thus when convolved with our instrument dispersion, the absorption lines are widened, therefore their impact on the DIB absorption will be limited. Then after selecting the best synthetic spectrum for an individual target, we subtract the synthetic spectrum from the observed absorption (containing both stellar and interstellar absorption features) to obtain the residual which predominantly consists of the interstellar absorption \citep{Montes95a,Montes95b}. We then fit a Gaussian function to this residual to obtain the equivalent width of the DIB (Fig.~\ref{dibsub}).

\subsection{Cool stars}
\begin{figure}
	\includegraphics[scale=0.38]{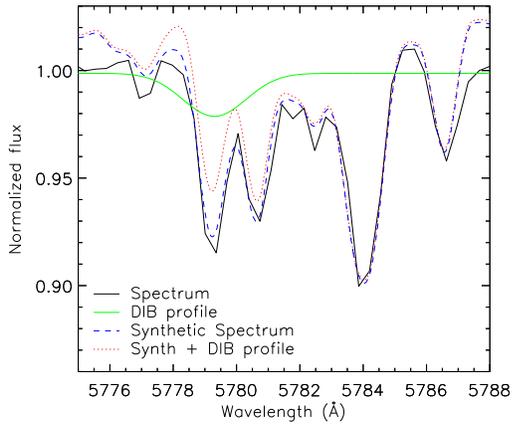}
	\caption{The spectrum of a cool star (black solid line). By changing the DIB profile parameters (thick solid line) and adding to synthetic spectrum (dotted line) the observed spectrum (dashed line) is reproduced.}
	\label{cooldib}
\end{figure}

The presence of numerous stellar lines in a cool star's spectrum (F, G, and K types) leads to an increased contamination in the interstellar absorptions and prevents an accurate determination of the EW, absorption depth, FWHM and the continuum. In these stars the \do has a two prong fork shape (see Fig.~\ref{cooldib}), which is caused by the presence of \mbox{Fe\,{\sc ii}}, \mbox{Mn\,{\sc i}}, \mbox{Si\,{\sc i}} (all near 5780.1 \AA) and \mbox{Cr\,{\sc i}} (5781.1 \AA). \cite{Chen13} studied the 6196, 6204.5 and 6283.8 \AA~DIBs and showed that, with the product of a synthetic stellar spectrum, a synthetic telluric transmission and an empirical model for the DIB absorption it would be possible to extract the spectrum of the interstellar feature. In our observations since we focus on \do and $\lambda5797$, the telluric absorptions are not considerable.

\cite{Chen13} show that DIB transmission profiles $D(\lambda)$ can be expressed as $D(\lambda) = D_{0}(\lambda)^{\beta}$ where $D_{0}(\lambda)$ is a reference profile derived by high signal to noise observations of early-type stars. Therefore by adjusting the $\beta$ parameters the proper profile for each DIB would be produced. According to \cite{Jacco09}, \do has a Gaussian profile:

\begin{equation}
D(\lambda) = a~\exp \left(  -\frac{(\lambda - b)^2}{2\sigma}  \right) + c
\label{eqd}
\end{equation}

After confirming the average \do DIB profile, according to Eq.~\ref{eqd}, in an iterative procedure we change the $ a $ (peak intensity), $ b $ (peak center) and $\sigma$ (peak width) for each wavelength ($\lambda$) to produce a new spectrum. Then we add this spectrum to the corresponding synthetic spectrum, and in each iteration according to Eq.~\ref{eqchi}, calculate the difference of the re-produced spectrum with the real observed absorption ($\chi^{2}$). Eventually, the best DIB profile estimation is the one with the smallest $\chi^{2}$ value.

\begin{equation}
\chi^{2} = \sum_{i=1}^{N}\left(\frac{\left( F(\lambda_{i}) - D(\lambda_{i}) \right)^{2}}{F_{{\rm err}}(\lambda_{i})^{2}}  \right)
\label{eqchi}
\end{equation}

In Fig.~\ref{cooldib}, it is clear that by adding the estimated DIB profile (thick solid line) to the synthetic spectrum (dashed line), the observed spectrum (solid line) is reproduced.

\begin{figure}
	\includegraphics[scale=0.38]{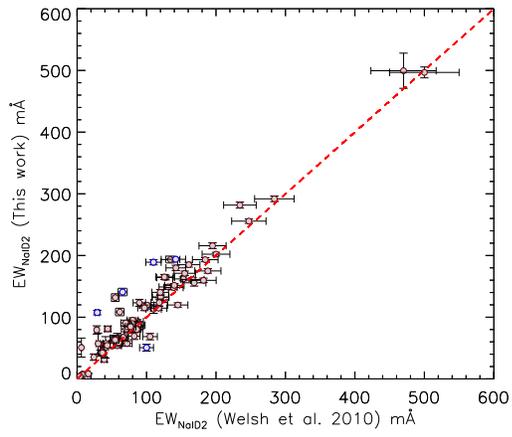}
	\caption{The comparison between \na measurements in this work and literature \citep{Welsh10}. The correlation coefficient in this plot is $c=0.97$. The most deviant objects are indicated with blue colour.}
	\label{comp}
\end{figure}

\begin{figure*}
	\includegraphics[scale=0.83]{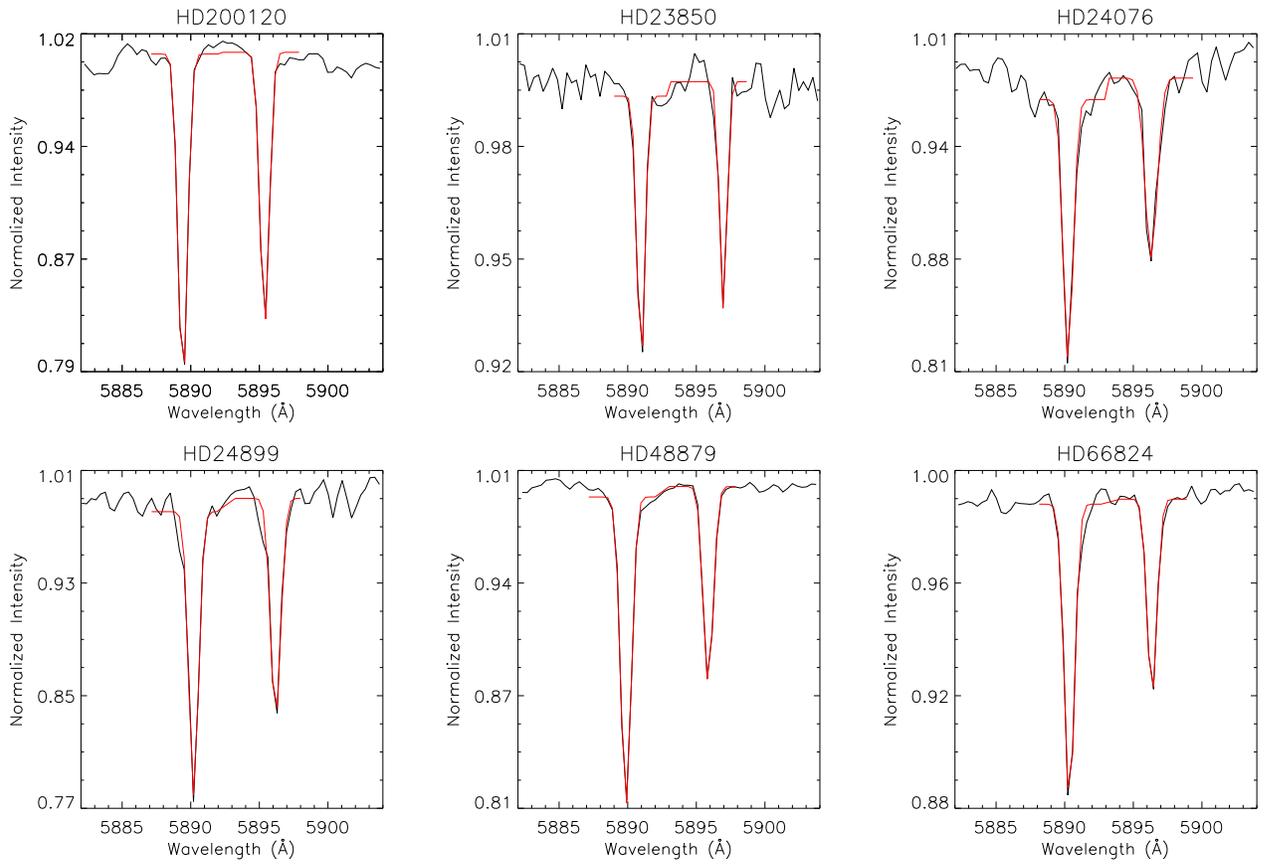}
	\caption{The most deviant objects from the \na comparison with the literature. All these targets are early-type stars and show strong \na doublet absorption.}
	\label{devs}
\end{figure*}

Fig.~\ref{comp} shows a comparison between the measured \na equivalent width in this work with \cite{Welsh10} measurements, with a correlation coefficient $c = 0.97$. The most deviant targets (indicated by different color)  in this plot are HD 200120, HD 23850, HD 24076, HD 24899, HD 48879 and HD 66824, all of which are early-type stars from B1 to A2 and are shown in Fig.~\ref{devs}. As is clear these targets have strong \na absorptions and hence the deviation is most probably caused by the difference in the spectral resolution of our observations ($R = 2000$) and Welsh ($R > 50000$) data. In addition the differences in the S/N ratio in the sets of measurements as well as the saturation level of both surveys may well play a role too.


\section{Equivalent widths measurements}
\label{results}
\begin{figure}
	\includegraphics[scale=0.43]{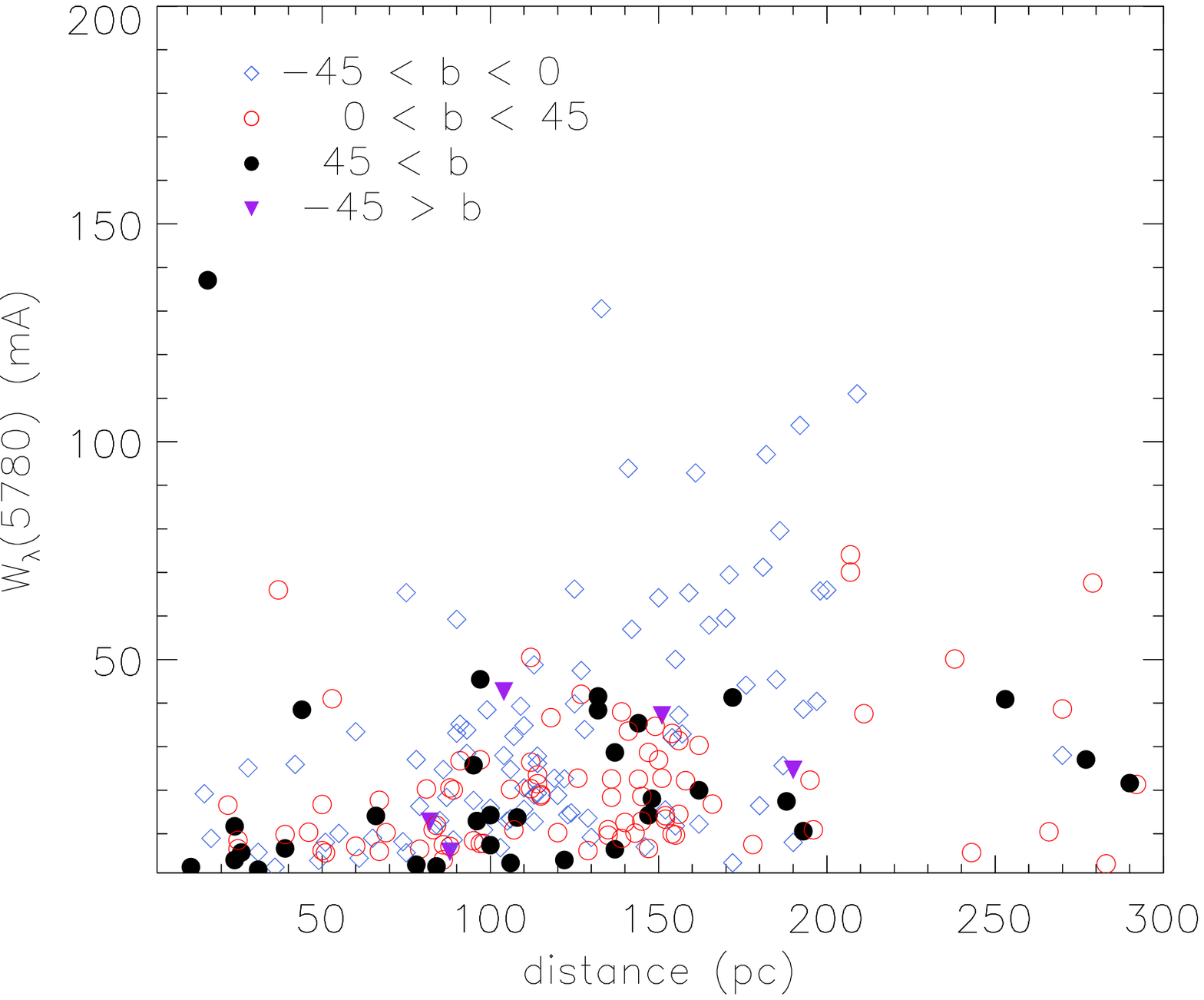}
	\caption{Equivalent width of \do versus distance from Sun for observed targets with distances $<$ 300 pc. Filled circles are for sightlines with b $> 45^{\circ}$, red open circles are for sightlines with b=0 to $45^{\circ}$, blue open diamonds are for sightlines with b=0 to $-45^{\circ}$ and filled triangles are for sightlines with b $< 45^{\circ}$.}
	\label{ew5780dis}
\end{figure}

In Fig.~\ref{ew5780dis} we show the total equivalent width of \do versus the distance from the Sun to $<300$ pc. \cite{Welsh10} according to a similar plot for \na D lines and Ca\,{\sc ii}, showed that the \na absorption up to $\sim$ 80 pc has very low values ($W_{\lambda}$(\na D2) $< 5$ m\AA), also within $\sim$ 100 pc there is little absorption for \ca ($W_{\lambda}$(\ca K) $< 15$ m\AA). Therefore, the volume of the Local Bubble is free of major dense interstellar neutral gas and only has some ionized warm cloudlets of Ca\,{\sc ii}. Also they show that beyond $\sim$ 80 pc and up to $\sim$ 100 pc a dense "wall" of neutral gas surrounds the Local Bubble and the value of $W_{\lambda}$(\ca K) increases slowly at this wall.

In contrast with neutral and ionized gas within the Local Bubble, in Fig.~\ref{ew5780dis} we show that, up to $\sim$ 85 pc there is a fair amount of DIB absorption ($W_{\lambda5780}>$15 m\AA). The presence of this absorption shows that within the Local Bubble, DIB cloudlets exist. However, there are several Galactic directions under the Galactic Plane (blue open diamond) with distances $>$100 pc which have strong DIB absorption. Also for Galactic directions which are located above the Galactic Plane (red open circles) there are some sightlines with strong DIB absorption at distances $>$200 pc extending into the Galactic Halo through the openings of the Local Chimney \citep{Lallement03}. The nearest stars with distances less than 80 pc that have anomalously high values of DIB absorption are HD 120136 (16 pc), HD 159332 (37 pc), HD 140436 (44 pc), HD 76756 (53 pc), HD 6658 (60 pc) and HD 218200 (75 pc).

\begin{figure}
	\includegraphics[scale=0.43]{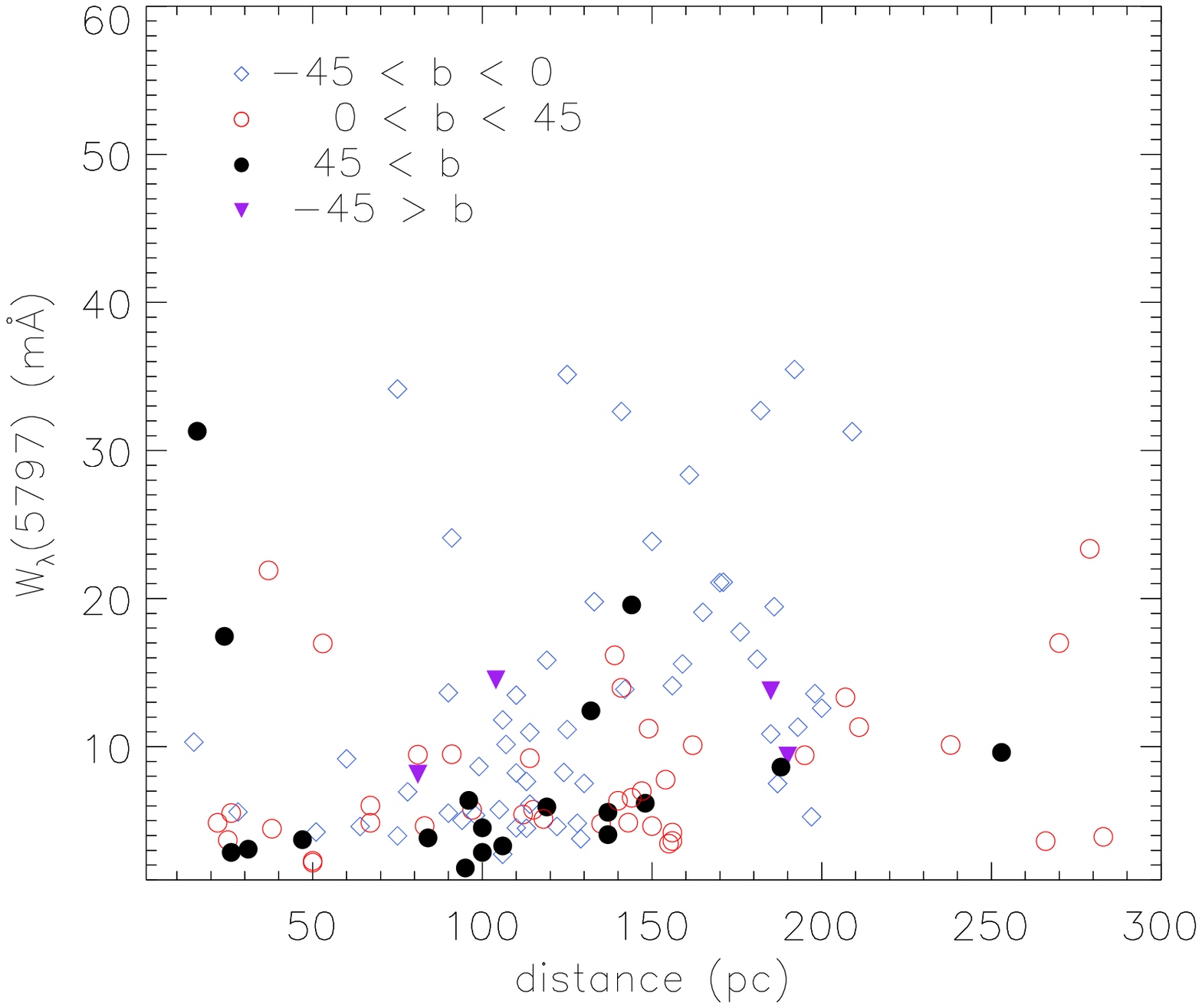}
	\caption{Equivalent width of \dt versus distance from Sun for observed targets with distances $<$ 300 pc. Filled circles are for sightlines with b $> 45^{\circ}$, red open circles are for sightlines with b=0 to $45^{\circ}$, blue open diamonds are for sightlines with b=0 to $-45^{\circ}$ and filled triangles are for sightlines with b $< 45^{\circ}$.}
	\label{ew5797dis}
\end{figure}

In Fig.~\ref{ew5797dis} we present the plot of \dt absorptions versus distance from the Sun, which shows that for several sightlines located within the Local Bubble up to $\sim$ 90 pc, the \dt absorption is strong ($W_{\lambda 5797} >$ 15 m\AA), and beyond the Local Bubble's wall, the majority of \dt equivalent widths are $W_{\lambda 5797} <$ 25 m\AA. Also for sightlines located in the Galactic Halo with distances $>$ 200 pc which are within the Local Chimney (red open circles) the \dt strength is $W_{\lambda 5797} <$ 15 m\AA. The two nearest stars within the Local Bubble with abnormal DIB strengths ($W_{\lambda 5797} >$ 30 m\AA) are: HD 120136 (16 pc) and HD 218200 (75 pc).

\begin{figure}
	\includegraphics[scale=0.94]{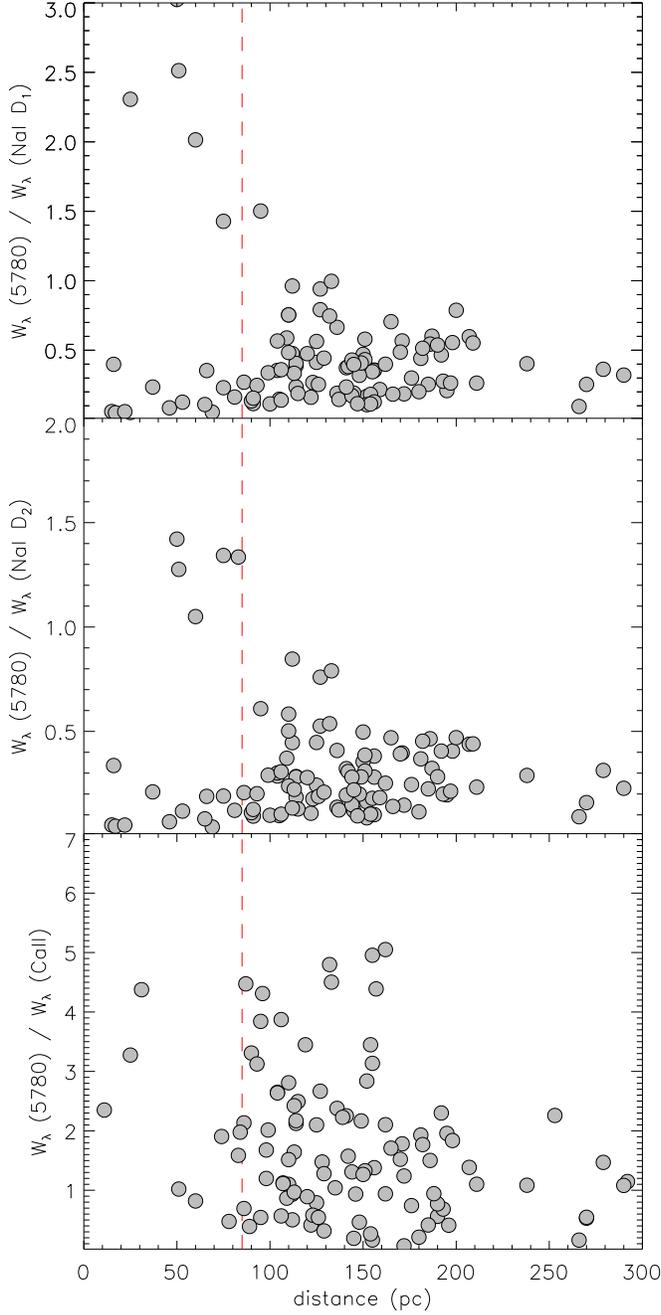}
	\caption{The equivalent width of \do to atomic lines ratio. The \textit{upper panel}: is $W_{\lambda}(5780)/W_{\lambda}($\na D$_{1})$ line. The \textit{middle panel}: is the $W_{\lambda}(5780)/W_{\lambda}($\na D$_{2})$ and the \textit{lower panel}: is the $W_{\lambda}(5780)/W_{\lambda}$(Ca\,{\sc ii}). The dashed line in all three panels is the d~80 pc boundary as a rough indication of the Local Bubble frontier.}
	\label{w80w97}
\end{figure}

In Fig.~\ref{w80w97} we show the \do equivalent width ratio to \na D lines and Ca\,{\sc ii}. Since the Local Bubble is depleted from \na D$_{1}$ and \na D$_{2}$ the $W_{\lambda}(5780)/W_{\lambda}$(\na D$_{1}$) and $W_{\lambda}(5780)/W_{\lambda}$(\na D$_{2}$) can also have high values below $\sim$ 80 pc, however beyond 80 pc the ratio falls. Meanwhile there is no significant change in $W_{\lambda}(5780)/W_{\lambda}$(Ca\,{\sc ii}) ratio within and beyond 80 pc. The difference between the $W_{\lambda}(5780)/W_{\lambda}$(\na D) ratio and $W_{\lambda}(5780)/W_{\lambda}$(Ca\,{\sc ii}) within and beyond the Local Bubble indicates that the \do is more associated with \na than Ca\,{\sc ii}. This will be discussed further in section \ref{caii}. The rapid \na ratio fall at the Local Bubble boundary is due to the presence of dense \na clouds around the Local Bubble which cause an increase in the equivalent width of both \na D lines, therefore the possibility of DIB existence increases too. Also, high values of \ca beyond 80 pc show that the \ca density does not follow the density of the \na \citep{Welsh10}.


\section{UV radiation and DIB ratio}
The UV radiation has a substantial effect on DIB carriers, for instance, \cite{Krelowski92} showed that the regions which are shielded from the UV radiation often have different ratios between equivalent widths of \do and $\lambda$5797. UV shielded sightlines are those line-of-sight which pass through the innermost regions of dense clouds, and following the observations of $\zeta$ Oph are named $\zeta$-type clouds. However, the sightlines passing through the outer layers of clouds are non-shielded from UV radiation, and following $\sigma$ Sco observations are named $\sigma$-type clouds \citep{Krelowski94}. \cite{Sonnentrucker97} concluded that the \do and \dt DIBs are separate gas-phase molecules. By comparing $W_{\lambda} / \bv$ as a function of \bv for each DIB, they found that \do reaches its maximum at the lowest value of \bv indicating that the carriers of this DIBs are the most resistant of DIBs to strong UV radiation \citep[so is the $\lambda$4429 DIB, see, e.g.,][]{vanLoon13}.

However, in more UV protected denser regions ($\zeta$-type), the \dt has a greater strength and the \do carrier is less efficiently ionized in these regions \citep{Vos11}. Therefore, in regions with high UV intensity the \do reaches its maximum abundance while the \dt carriers are ionized and then destroyed \citep{Cami97}. As a result, this suggests that the formation of the \do carriers requires UV photons. Also since the \dt has a positive correlation with the overall slope of the extinction curve and anti-correlates with the UV radiation, the formation of the \dt carrier would benefit from the UV shielding \citep{Megier05}. Consequently, the abundance ratio between \do and $\lambda5797$, to some extent traces the ionization state of the DIB carriers \citep{Vos11}.

The original definition of these two types of clouds is based on the ratio of central depths of the \dt and \do DIB absorptions ($A5797/A5780=0.4$), which corresponds to the equivalent width ratio, $W_{\lambda}(5797) / W_{\lambda}(5780) \simeq 0.3$. Since there is a smooth transition, the exact boundary is not very important \citep{Kos13, Vos11}.

\begin{figure}
	\includegraphics[scale=0.47]{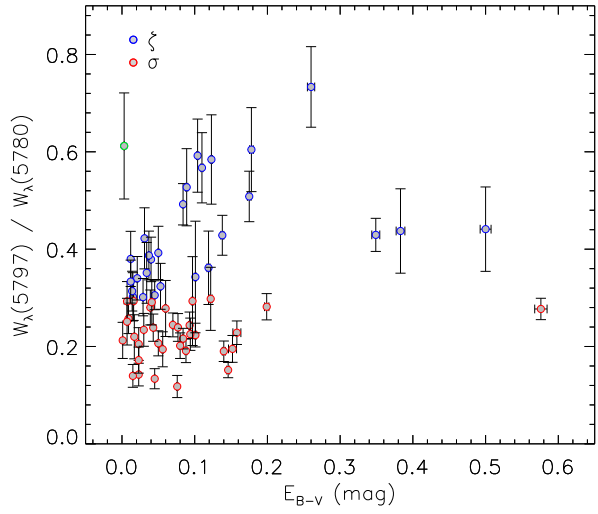}
	\caption{The $ W_{\lambda}(5797)/W_{\lambda}(5780)$ ratio plotted against E\,{\textsubscript {B-V}}. The distribution peaks at an \bv of $\sim$ 0.14 mag, indicating the optimal condition for \do formation.}
	\label{UVwRatio}
\end{figure}

In Fig.~\ref{UVwRatio} we show the $W_{\lambda}(5797) / W_{\lambda}(5780)$ ratio for observed $\zeta$-type and $\sigma$-type clouds versus reddening. For sightlines with very low $\bv < 0.1$ mag, the $\sigma$-type clouds are sufficiently abundant, which shows that the \do carrier is more abundant than the \dt carrier. The peak around \bv $\sim 0.14$ mag corresponds to sufficient shielding so there is the most optimal condition for the existence of the \dt carrier and instead there is insufficient UV flux to transform the \do carrier into its ionic form \citep{Vos11}. For sightlines with \bv$> 0.25$ mag (with sufficient UV protection) the condition for formation of the \dt carriers is more favorable than in the $\sigma$ sightlines. But after \bv$> 0.25$ mag, there is no more increase in $\zeta$-type clouds, because either the \dt carrier is depleted at high densities and/or these are simply caused by superpositions of clouds with individually lower reddening. The target with a different color (green) is HD 222602, an early-type star on the main sequence with a very low color index ($B-V = +0.1$ mag). This star exhibits very strong \do and \dt absorption in a high-quality spectrum.

\begin{figure}
	\includegraphics[scale=0.38]{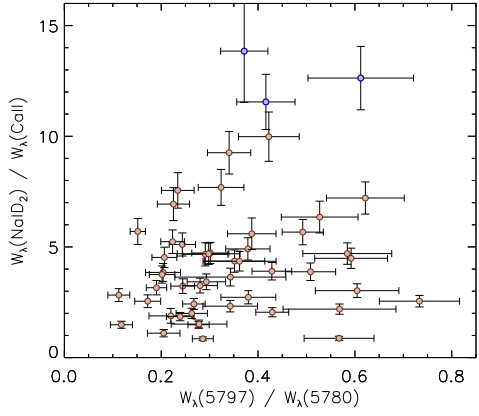}
	\caption{The $ W_{\lambda}(5797)/W_{\lambda}(5780) $ ratio plotted against $ W_{\lambda}($\na D$_{2})/W_{\lambda}($Ca \,{\sc ii}). The two most scattered targets are identified with blue color.}
	\label{ratio}
\end{figure}

In Fig.~\ref{ratio} we plot the $W_{\lambda}(5797) / W_{\lambda}(5780)$ versus $W_{\lambda}($Na\,{\sc i} D$_{2})/W_{\lambda}($Ca\,{\sc ii}$)$, which shows that when the $W_{\lambda}(5797) / W_{\lambda}(5780)$ increases the $W_{\lambda}($Na\,{\sc i} D$_{2})/W_{\lambda}($Ca\,{\sc ii}$)$ increases too. In other words, the \dt carrier is associated with regions with higher \na absorption, probably in the dense cores of clouds. Also in environments where the \ca is strong, the DIBs ratio is weak, meaning that the \do strength is high or/and the \dt is significantly weak. The most deviant measurements are for HD 151525, HD 164577 and HD 25642 which all have a very good quality and high S/N ratio, but all sightlines have a high $W_{\lambda}($Na\,{\sc i} D$_{2})/W_{\lambda}($Ca\,{\sc ii}$)$ ratioes.


\section{Correlations of DIB features}
\subsection{Correlation between DIBs and extinction}
The DIBs are correlated well with visual reddening \bv and among all known DIBs, the \do has the strongest correlation \citep{Herbig95}. \cite{Megier05} studied the DIB correlations with the slope of the extinction curve between the far-UV and other parts of the extinction curve, and also the DIB correlation with the 2175 \AA~bump strength. They found that amongst all DIBs, the \do has the strongest correlation with the bump, which could be interpreted as both the bump and the DIB carriers being composed of similar materials. There is a loose correlation between the 2175 \AA~bump and the extinction excess in the far-UV, which suggests that the absorbers might be carbonaceous molecules such as PAHs, particularly ionized species. Also other identifications have been proposed for DIB carriers such as fullerenes\footnote{Fullerenes are spherical or elliptical shells formed with hexagonal and pentagonal aromatic cycles like C$_{60}$}.

The DIB strengths could reflect changes in grain properties, for instance, in some directions (e.g., Sco$ - $Oph) the slope of the reddening law ($R = A_{V}/E_{B-V}$) departs from the standard value\footnote{the overall value is $R=3.1$ but in Sco$ - $Oph varies between $R=4-4.4$ \citep{Cardelli89}}. Such variations in R are believed to be due to differences in the mean size of the grains in those sightlines. However, \cite{Herbig93} showed that DIB strengths per unit color excess do not follow the R variations, so this provides a reason to doubt whether the grains themselves can be hosts of the carriers.

To investigate the DIB correlation with visual reddening, we divide our data into two different groups. All correlation coefficients are computed for sightlines for which the star resides inside the Local Bubble and those in which the target star is outside the Local Bubble. In this study the Johnson UBV data were taken from \cite{Kharchenko04}, \cite{Gontcharov06} and \cite{Kharchenko07}, and the stellar types were selected from \cite{Hoffleit95}. In addition, the (B$ - $V)$_{0}$ intrinsic colors were taken from \cite{Fitzgerald68}. For extracting reddening errors we follow the method of \cite{Friedman11} which results in \bv errors of 0.03 mag.

\begin{figure}
	\includegraphics[scale=0.56]{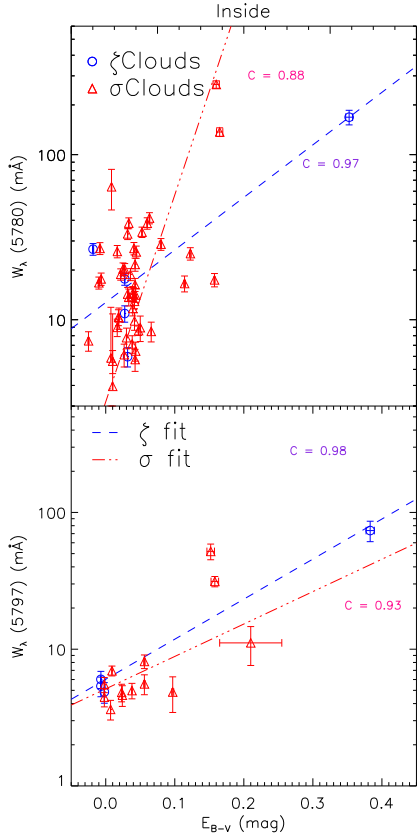}
	\caption{Equivalent width of \do (upper panel) and \dt (lower panel) versus reddening \bv for sightlines located inside the Local Bubble. All sightlines are divided in two different $\sigma$ and $\zeta$ sightlines. The triangles are $\sigma$ clouds and circles are $\zeta$ clouds. Also the dashed line is the best-fit line to $\zeta$ sightlines and dot-dashed line is the best-fit to $\sigma$ clouds. The correlation coefficients are given by c.}
	\label{UVin}
\end{figure}

In Fig.~\ref{UVin} we show the \do and \dt correlation with \bv in $\sigma$-type and $\zeta$-type clouds inside the Local Bubble. This classification reduces the scatter and improves the relation between DIB strength and reddening. As Fig.~\ref{UVin} shows, the \do and reddening show a stronger correlation for $\zeta$ sightlines compared to $\sigma$ sightlines. Although the scatter at the lower end of the relations is large, it appears that the slope of the relation for the $\sigma$-type clouds is steeper than the same for $\zeta$-type clouds for \do, indicating the \do carriers are located in low density regions of clouds. In the presence of UV radiation, the rate of molecule dissociation increases and the ionic compounds separate or split into smaller particles; also, the clouds within $\sigma$ environments may have the denser, dustier cores that would result in an increased reddening. \cite{Kos13} showed that the linear relation between the extinction and \dt EW for $\zeta$-type clouds is significantly steeper than the relation for $\sigma$ sightlines, which indicates that the \dt is associated with denser parts of clouds. As the lower panel of Fig.~\ref{UVin} shows, the best fit line through \dt and \bv is steeper in $\zeta$-type clouds, which implies that the \dt is located in denser parts of clouds, protected from UV radiation. Stronger correlation between the \do and \dt and the reddening for $\zeta$-type clouds, in both panels, is natural as there may be more dust in dense regions of the clouds.

\begin{figure}
	\includegraphics[scale=0.56]{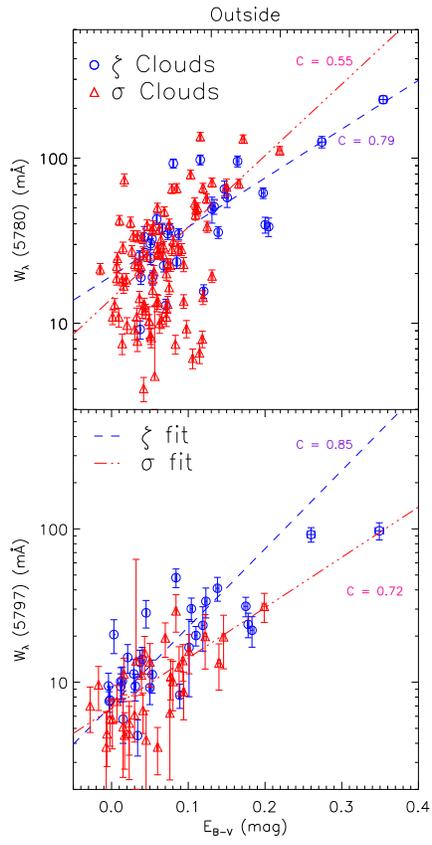}
	\caption{As Fig.~\ref{UVin} but for outside the Local Bubble.}
	\label{UVout}
\end{figure}

In Fig.~\ref{UVout} we show the $\zeta$-type and $\sigma$-type clouds outside the Local Bubble for both \do and $\lambda$5797. Both correlations for $\sigma$ and $\zeta$-type clouds are high, but the $\zeta$-types seem to have stronger correlations than the $\sigma$-type clouds. But based on these data this does not appear to be convincing. In the upper panel, the $\sigma$-type clouds show a steeper linear relation between \bv and \do which indicates that \do grows in the edges of clouds. Also, in the lower panel, the \dt and \bv correlation outside the Local Bubble is steeper for $\zeta$-type clouds, caused by the presence of the \dt DIBs in the denser parts of clouds that are protected from UV radiation.

Within the Local Bubble HD 214994, HD 204862 and HD 140436 are most deviant from the best fit line. These stars have strong \do absorptions and are situated within a region with low reddening. Also, HD 18055 located beyond the Local Bubble has strong \do absorption while it resides in a low dust region. Similarly, the most deviant targets in \dt vs \bv outside the Local Bubble are HD 140728, HD 18484, HD 20809 and HD 23016 which have strong \dt absorptions with high S/N ratio.

\subsection{Correlation between DIBs and \na}
\cite{Herbig93} showed that the log W(5780) and log W(5797) follow a linear dependence on log N(\mbox{Na\,{\sc i}}) but with a lot of scatter. He noted that the existence of these correlations with neutral atoms does not imply that the DIB carriers are neutral molecules, since if approximately the degree of first ionization was constant, equally correlations would probably be found for the ions.

\begin{figure*}
	\includegraphics[scale = 0.77]{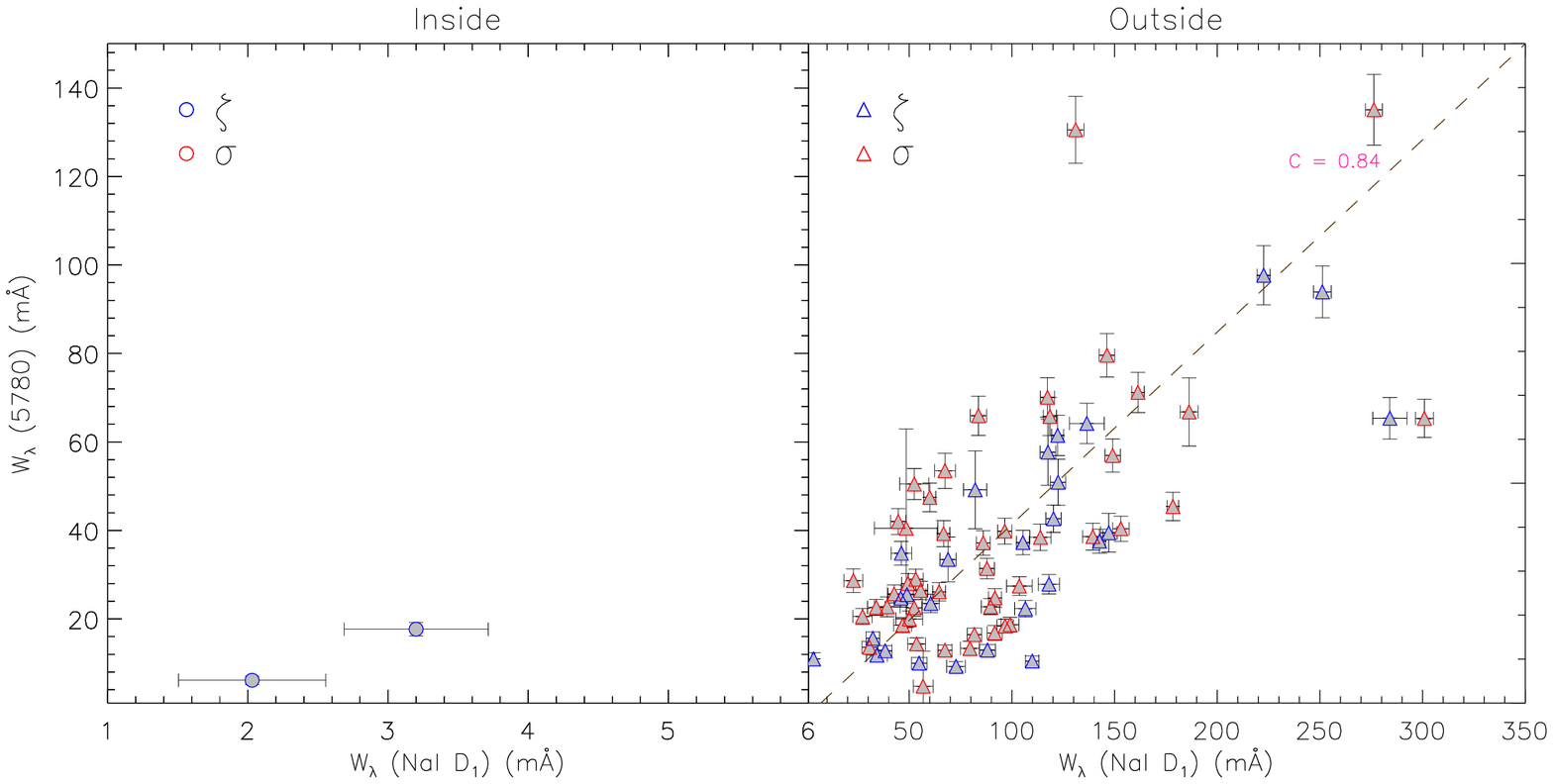}
	\caption{Correlation between the \do and \na D$_{1}$. \textit{Left panel}: within the Local Bubble and \textit{Right panel}: outside of the Local Bubble. The dashed line is the best-fit line and the correlation coefficient is given by c.}
	\label{NaID1vs5780}
\end{figure*}

\begin{figure*}
	\includegraphics[scale = 0.77]{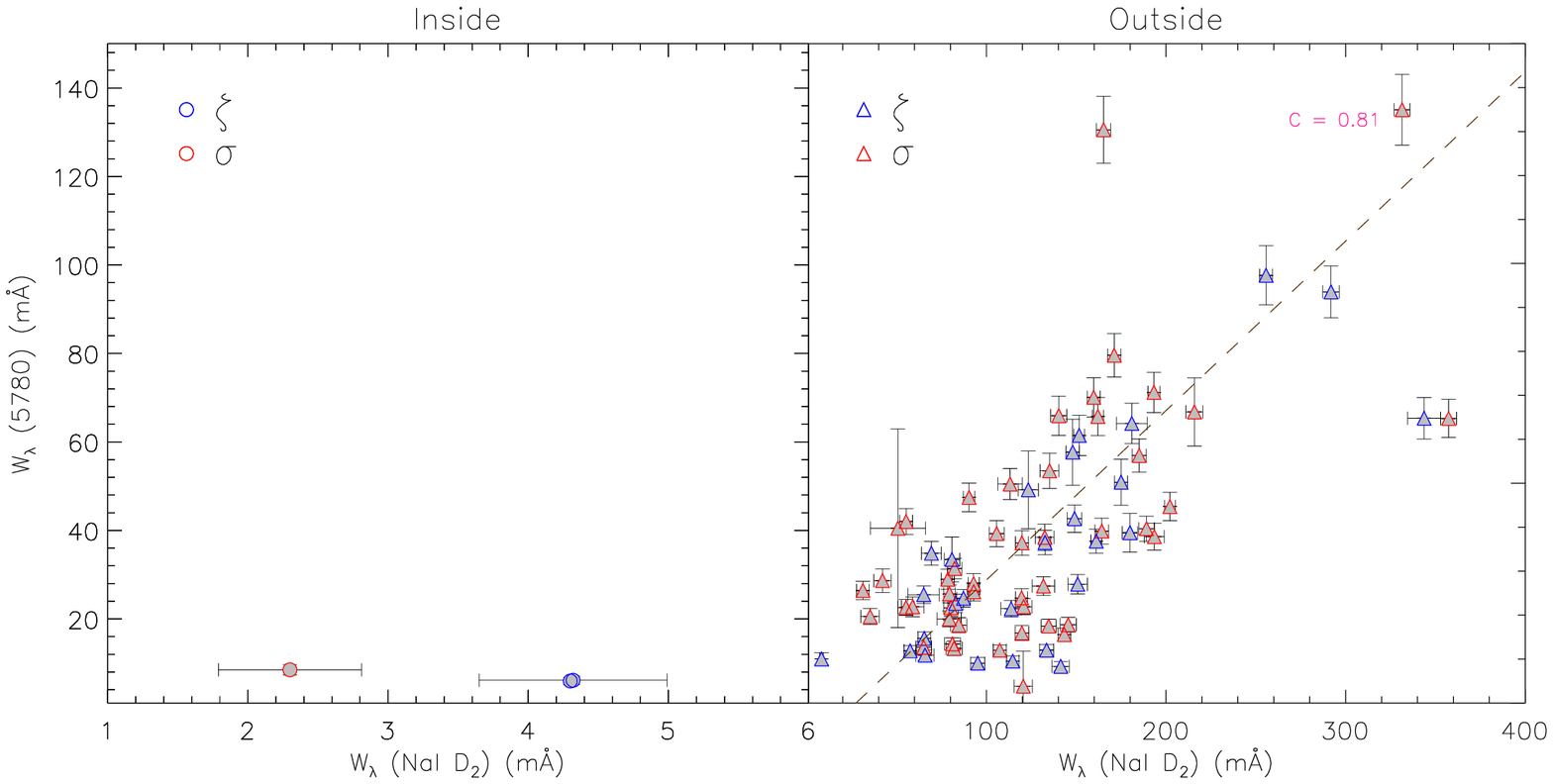}
	\caption{Correlation between the \do and \na D$_{2}$. \textit{Left panel}: within the Local Bubble and \textit{Right panel}: outside of the Local Bubble. The dashed line is the best-fit line and the correlation coefficient is given by c.}
	\label{NaID2vs5780}
\end{figure*}

In Fig.~\ref{NaID1vs5780} we plot the \do correlation with neutral \na D$_{1}$\footnote{D refers to all the transitions between the ground state and the first excited state of the Na$^0$ atoms and the D$_{1}$ and D$_{2}$ lines correspond to the fine-structure splitting of the excited states respectively with 5895.9 \AA~and 5889.9 \AA.}. As is shown in the figure, the targets with measurable \do values within the Local Bubble are very low in numbers, this is because of depletion of \na D$_{1}$ from within the Local Bubble. Therefore it is hard to speak about correlations and linear relations. The \do and \na D$_{1}$ correlation for the outside sightlines is clear ($c = 0.84$). In general, the existence of a common correlation between DIBs and \na D$_{1}$ in dense clouds does not imply that DIB carriers have physical similarity to Na, rather it shows that in each environment where the density increases the DIB strength increases too. Likewise, we find a correlation between \do and \na D$_{2}$ outside the Local Bubble (Fig.~\ref{NaID2vs5780}). Outside the Local Bubble, HD 22091, HD 218200 and HD 27820 are most deviant from the best fit line. In these sightlines, the \na absorptions are strong but the \do features are weak.

\begin{figure*}
	\includegraphics[scale = 0.77]{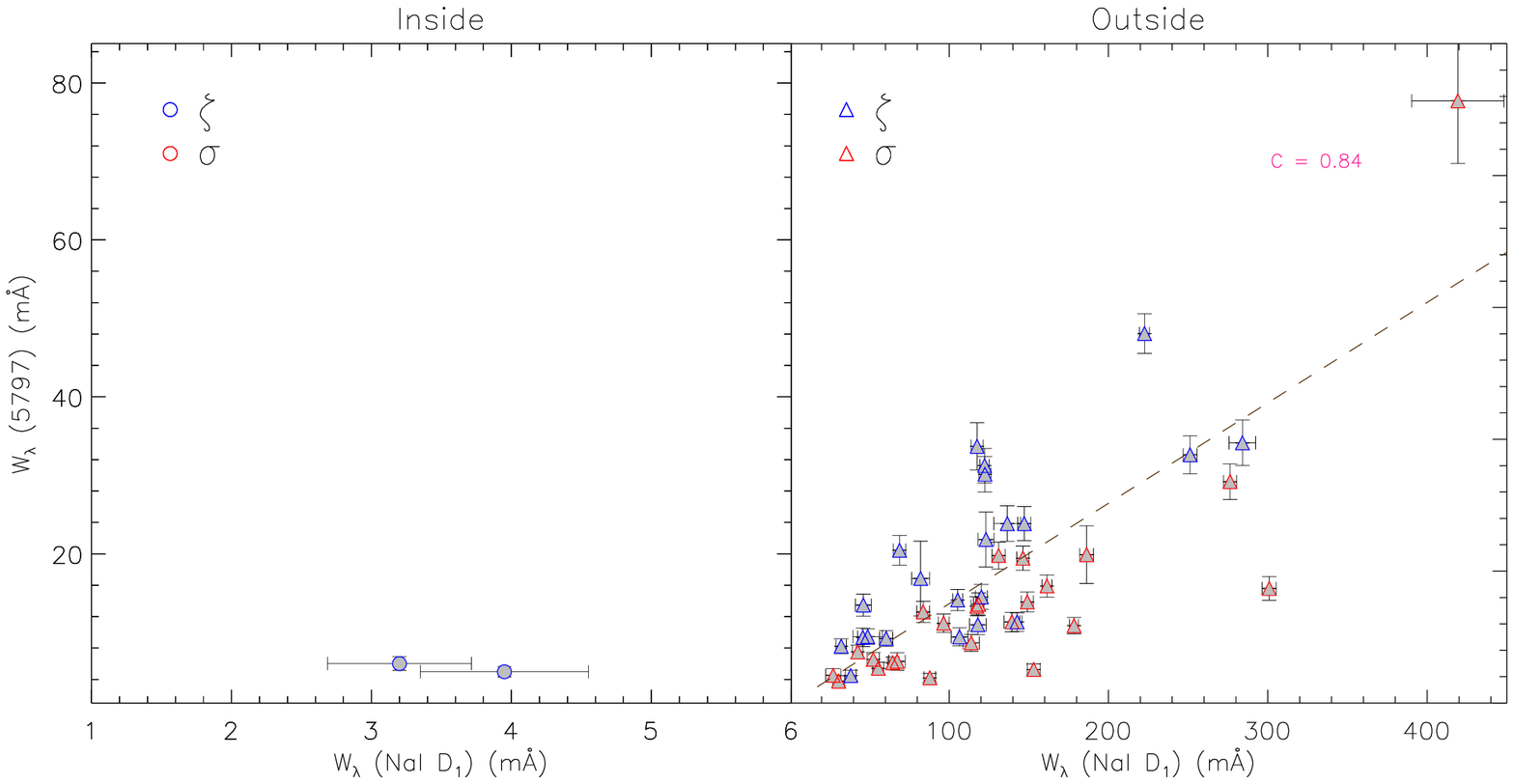}
	\caption{Correlation between the \dt and \na D$_{1}$. \textit{Left panel}: within the Local Bubble and \textit{Right panel}: outside of the Local Bubble. The dashed line is the best-fit line and the correlation coefficient is given by c.}
	\label{NaID1vs5797}
\end{figure*}

\begin{figure*}
	\includegraphics[scale = 0.77]{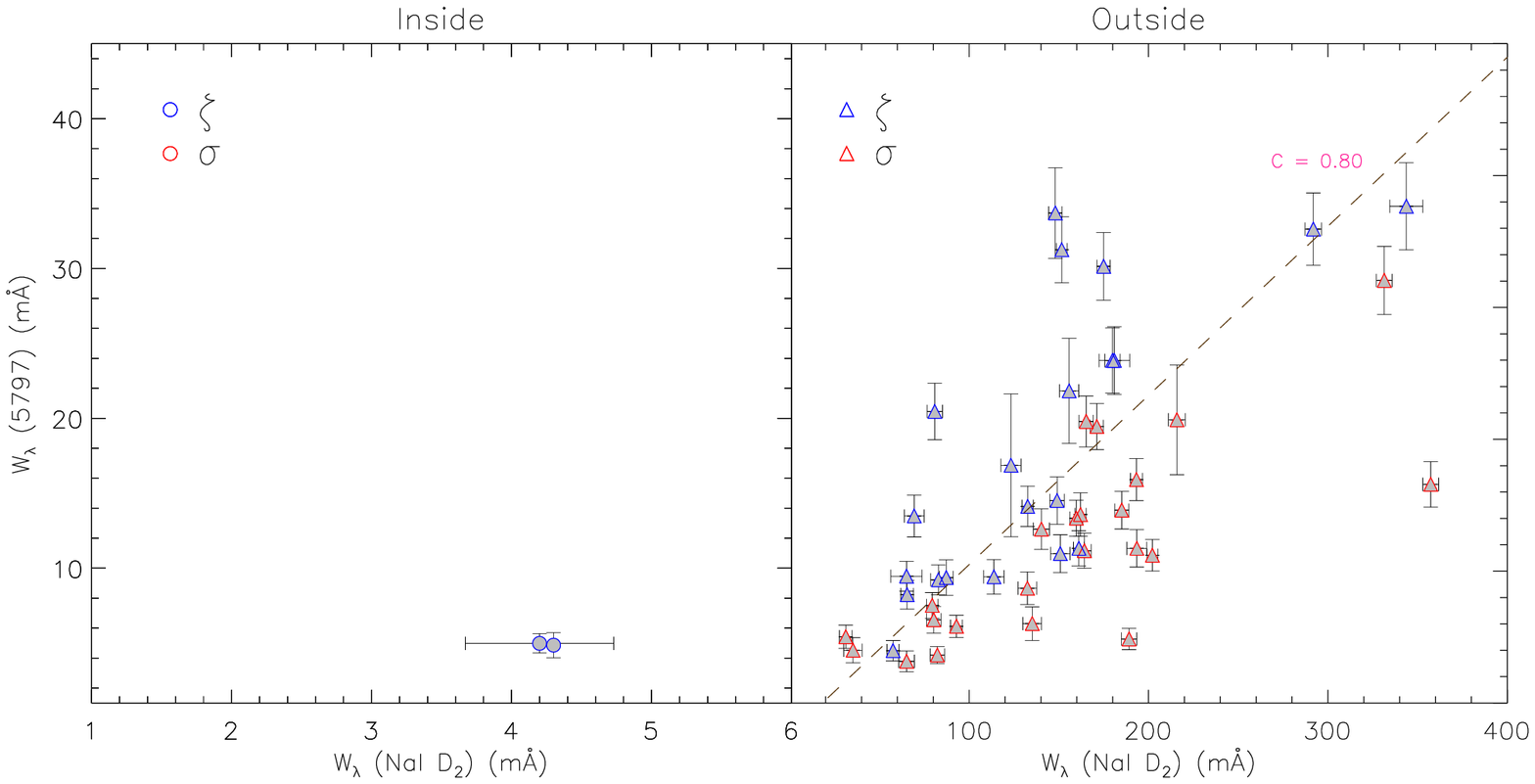}
	\caption{Correlation between the \dt and \na D$_{2}$. \textit{Left panel}: within the Local Bubble and \textit{Right panel}: outside of the Local Bubble. The dashed line is the best-fit line and the correlation coefficient is given by c.}
	\label{NaID2vs5797}
\end{figure*}

In Fig.~\ref{NaID1vs5797} we show the \dt strength versus \na D$_{1}$. Outside the Local Bubble, \dt has a common correlation with the D$_{1}$ line ($c = 0.84$). If we consider \dt with \na D$_{2}$ there is a similar manner of correlation inside the Local Bubble (Fig.~\ref{NaID2vs5797}). Despite the lack of data inside the Local Bubble, they are all located in $\zeta$-type clouds and hence are in the denser parts of clouds. Again, outside the Local Bubble there is a common correlation between \dt and \na D$_{2}$ ($c = 0.80$).

HD 42088, HD 20809 and HD 22091 are deviant from the \dt and \na D$_{1}$ linear fit outside the Local Bubble. Two first targets are located above the best fit line, but HD 22091 target is seen through dense \na clouds, while its \dt absorptions is weak. Likewise, HD 23016, HD 25940 and HD 22091 are deviant from the linear fit through \dt and \na D$_{2}$, such that the HD 23016 and HD 25940 are above the best fit line, and the HD 22091 have strong \na absorptions and weak $\lambda5797$.

\subsection{Correlation between DIBs and \ca}
\label{caii}
We use a \ca K-line (3393 \AA) catalog \citep{Welsh10} to investigate the existence of possible correlations between the DIBs and Ca$^{+}$. This could provide information on the energy levels that are responsible for the existence, excitation and ionization of the carriers \citep{Galazutdinov04}. Some nearby early A-type stars have circumstellar \ca lines, caused by the debris disks around these young stars. These \ca lines may easily be confused with interstellar \ca absorption \citep{Montgomery12}. Because circumstellar \ca can only be identified by repeated observation and not in a single observation, some of the \ca measurements reported by \cite{Welsh10} may have been affected by circumstellar absorption.

\begin{figure*}
	\includegraphics[scale = 0.77]{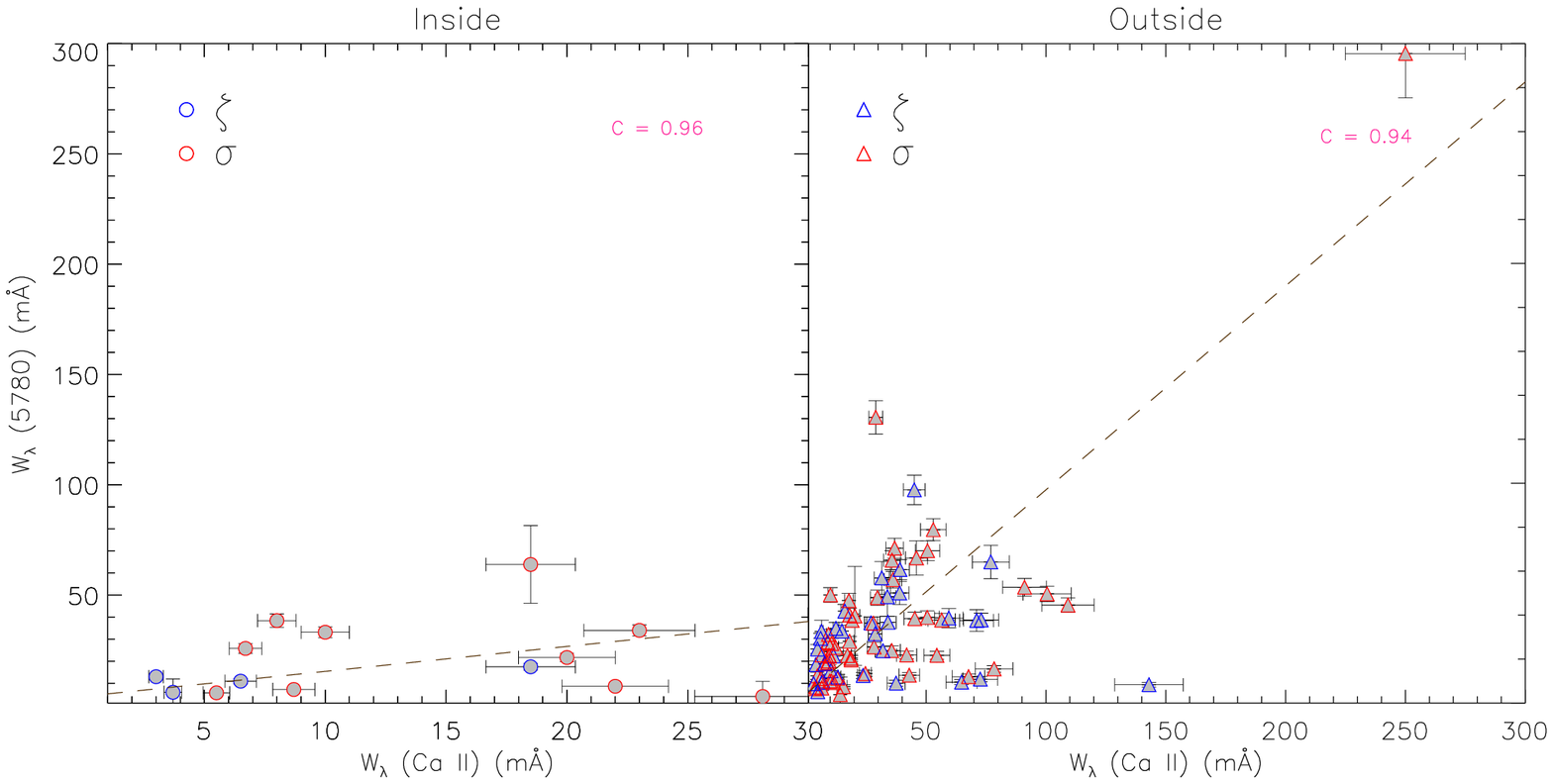}
	\caption{Correlation between the \do and Ca\,{\sc ii}. \textit{Left panel}: within the Local Bubble and \textit{Right panel}: outside of the Local Bubble. The dashed line is the best-fit line and the correlation coefficient is given by c.}
	\label{CaIIvs5780}
\end{figure*}

\begin{figure*}
	\includegraphics[scale = 0.77]{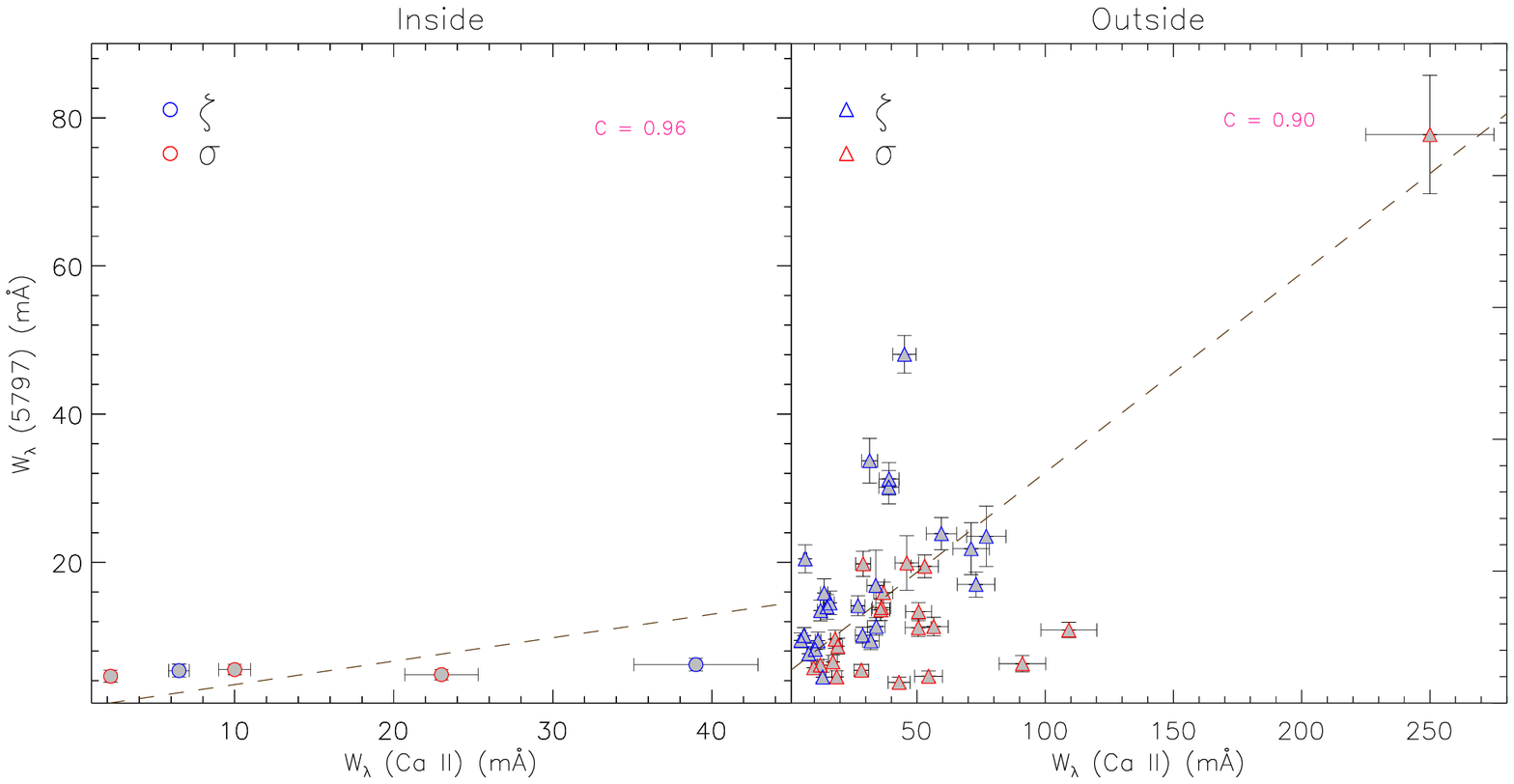}
	\caption{Correlation between the \dt and Ca\,{\sc ii}. \textit{Left panel}: within the Local Bubble and \textit{Right panel}: outside of the Local Bubble. The dashed line is the best-fit line and the correlation coefficient is given by c.}
	\label{CaIIvs5797}
\end{figure*}

As is shown in Fig.~\ref{CaIIvs5780}, the \do and \ca correlation outside the Local Bubble is obvious, suggesting that this diffuse band is found preferentially in the same regions where \ca forms. There is a hint of a similar trend inside the Local Bubble but the scatter is large. The \dt correlates well with \ca outside the Local bubble, and a similar trend can be seen inside the Local Bubble (Fig.~\ref{CaIIvs5797}). The correlation between \ca with \do is stronger than its correlation with $\lambda5797$, suggesting that \do is more associated with regions where Ca$^{+}$ is more abundant. Given that the \ca appears in more radiant regions, the \do macromolecule may have an electrically charged carrier \citep [for more discusion see][]{Jacco09}.

Inside the Local Bubble, absorptions associated with $\zeta$-type clouds have small \ca values ($\le 5$ m\AA), for both DIBs. This shows that the dense regions of clouds are cooler and thus depleted from \ca atoms. Also within the Local Bubble, in the $\sigma$-type clouds, the \ca is more intense and the \do intensity is obviously higher than the \dt values. This suggests that within the Local Bubble the $\sigma$-type clouds are dominant in the warmer mediums. 

The most deviant targets from the best fit line of \ca and \do outside the Local Bubble are HD 216057, HD 77770, HD 220599 and 40724, which have strong \ca absorption but relatively weak \do feature. Also in the direction of HD 26912, the \do absorption is strong but the \ca located in this sightline is weak. For \dt and \ca outside the Local Bubble, the most deviant targets are HD 216057 and HD 166182 which have strong \ca absorptions and weak \dt feature. But the HD 20809 have strong \dt absorption but a weak \ca feature.

\subsection{Correlation between DIBs}
\label{dibcor}
\begin{figure*}
	\includegraphics[scale = 0.77]{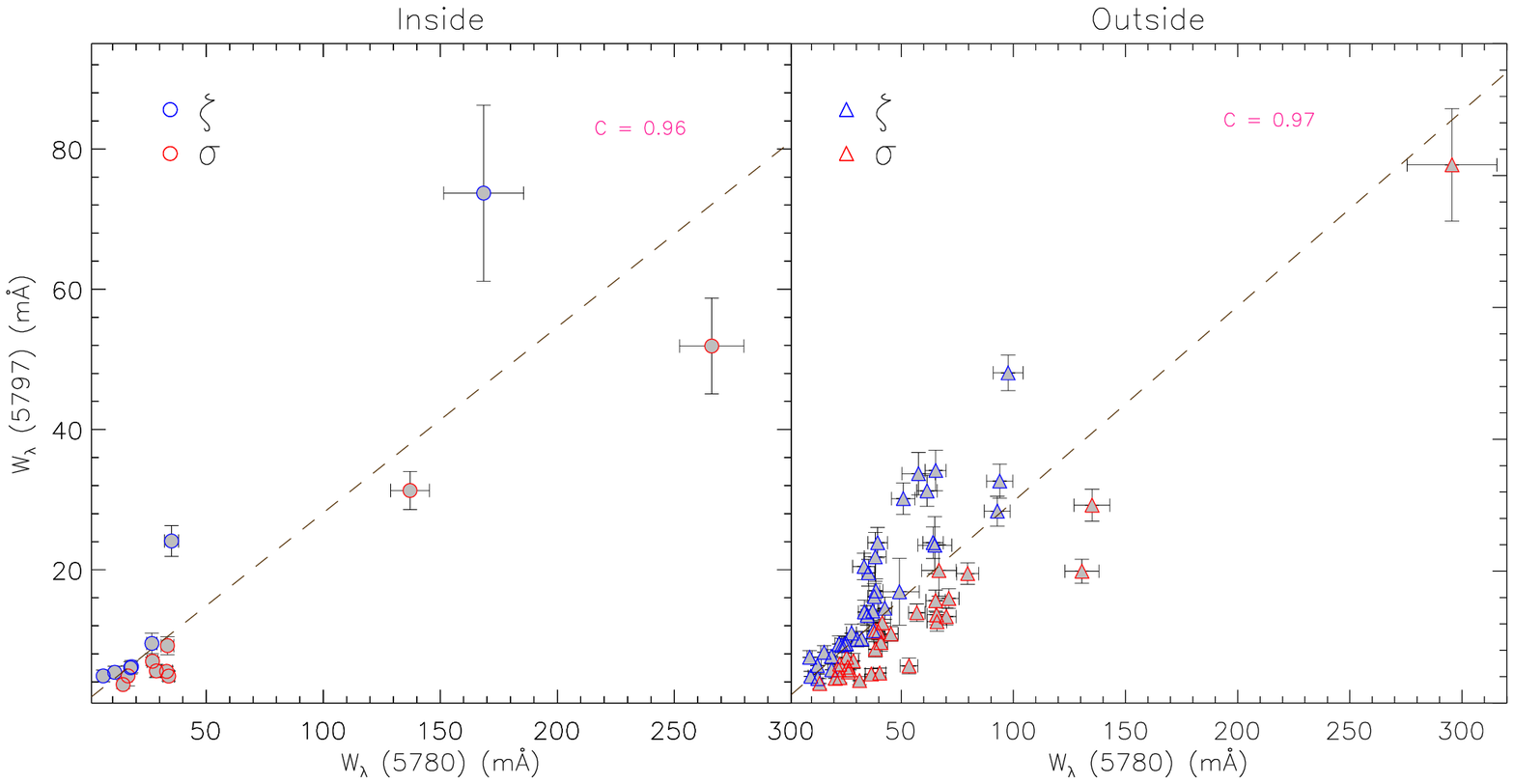}
	\caption{Correlation between the \do and $\lambda5797$. \textit{Left panel}: within the Local Bubble and \textit{Right panel}: outside of the Local Bubble. The dashed line is the best-fit line and the correlation coefficient is given by c.}
	\label{5780vs5797}
\end{figure*}

As shown in Fig.~\ref{5780vs5797} there is a fairly clear correlation between \do and \dt both outside and inside the Local Bubble. This suggests both DIB carriers form, or are existed, in environments which are generally found to co-exist. The scatter is larger inside the Local Bubble than outside, reflecting the more diffuse ISM within the Local Bubble and the correspondingly greater sensitivity to small variations in cloud structure.

If linear fitting between equivalent widths of two species does not pass through the origin, this may be evidence of a threshold effect such that a substantial amount of one kind must be produced before the other. \cite{Friedman11} showed that among all DIB correlations with $\lambda5780$, only $\lambda$6613.6 and $\lambda$6283 have a linear fit that does not pass through the origin. They concluded that the \do DIB must be produced before the $\lambda$6613.6 begins to form, and $\lambda$6283 appears before $\lambda5780$. In our observations, as shown in Table~\ref{t1a}, similarly we find that the intercept of the linear fitting for the correlation of \do and \dt passes through the origin. This shows that both \do and \dt originate from one source, in and around the Local Bubble.

\section{Conclusions}
\label{discus}
High signal to noise spectra of 432 hot stars within a distance of $\sim$ 200 pc were obtained, with the aim to study the Local Bubble and its surroundings. All observations were taken in the Northern hemisphere, complementing the Southern survey (Paper I). We investigate the spatial variation of the \do and \dt strengths within and beyond the Local Bubble. We also compare the \do and \dt absorptions with interstellar reddening, Na\,{\sc i} and Ca\,{\sc ii}. Our main conclusions are summarized as follows:
\begin{itemize}
	\item The $W_{\lambda}(5797) / W_{\lambda}(5780)$ ratio versus \bv indicates that the $\sigma$-type clouds are distributed around very low \bv$< 0.1$ mag. Meanwhile since the \do is more abundant in the $\sigma$-type clouds, therefore the peak shows that the \do is tend to be present in irradiated regions with low UV shielding. Also, the $\zeta$ peak is found around \bv $\sim 0.14$ mag which indicates sufficient UV shielding. Since the \dt is more prevalent in the $\zeta$-type clouds, the peak shows that the \dt carriers grow in regions with sufficient shielding where they are protected from the background UV radiation.

	\item For both inside and outside the Local Bubble, the \do and reddening relation is steeper for sigma sightlines than for $\zeta$ sightlines. This suggests that in $\zeta$ sightlines the denser, dustier cores of clouds become more dominant, increasing the reddening but not leading to a similar increase in \do absorption as this DIB carrier resides in the more diffuse gas. In contrast, the \dt and reddening relation is steeper in $\zeta$ sightlines because the \dt carrier is more closely associated with denser parts of clouds where it is shielded from UV radiation.

	\item The $W_{\lambda}(5797) / W_{\lambda}(5780)$ ratio increases with $W_{\lambda}($Na\,{\sc i} D$_{2})/W_{\lambda}($Ca\,{\sc ii}$)$ ratio. It is fairly established that, in the dense core of clouds the \na and \dt is more abundant and as the cloud becomes less dense toward the outer regions of the cloud, the \na becomes weaker and the \do DIB appears. Also the outer regions are subject to greater irradiations and the \na disappears and \ca starts to dominate \citep{Jacco09}. Therefore when the $W_{\lambda}(5797) / W_{\lambda}(5780)$ ratio increases, it means that we are increasingly dominated by the cores of clouds that are more shielded from background UV, so the $W_{\lambda}($Na\,{\sc i} D$_{2})/W_{\lambda}($Ca\,{\sc ii}$)$ ratio increases too.
\end{itemize}

	For outside the Local bubble the \do and \dt are well correlated with the \na doublet lines. But the \do correlation is stronger than the \dt correlation, therefore the sightlines outside the Local Bubble are translucent and hence the $\sigma$ behavior would be expected to dominate.
	
	The \do and \dt are well correlated with \ca outside the Local Bubble, suggesting that these diffuse bands are found preferentially in the same areas where \ca forms. There is a hint of a similar trend inside the Local Bubble but with a greater scatter. \cite{Jacco09} mentioned that the \ca trace the warm medium, therefore maybe the higher correlation value for inside sightlines caused by the high temperature of Local Bubble.

\acknowledgments
We wish to thank the Iranian National Observatory (INO) and School of Astronomy at IPM for facilitating and supporting this project. The observing time allocated to this project was provided by the INO. We also wish to thank the ING staff, scientific, technical and admin, for their support. Some of the research visits related to this project have been supported by the Royal Society International Exchange Scheme. Farhang Habibi, Ehsan Kourkchi, Alireza Molaeinezhad, Saeed Tavasoli, Sara Rezaei and Maryam Saberi contributed to this project in observing runs as part of the INO training program with the INT. Liam Hardy contributed to this project as a substitute observer. Furthermore, we thank the referee for a very thorough reading of the manuscript and many helpful suggestions that improve this paper.

{\it Facilities:} \facility{Isaac Newton Telescope (INT)}.

\clearpage

\clearpage

\clearpage

\clearpage

\clearpage

\clearpage

\clearpage

\clearpage

\clearpage

\clearpage

\clearpage

\clearpage

\begin{table}
\small
\centering
	\caption{The correlations of \do and \dt with each other and with \na D$_{1}$, \na D$_{2}$ and \ca equivalent widths. A distinction is made between sightlines inside and outside of the Local Bubble.}
	\begin{tabular}{@{}ccccccccc@{}}\toprule
		~ & \multicolumn {4} {c} {Inside the Local Bubble} & \multicolumn {3} {c} {Outside the Local Bubble} \\
		\cmidrule{2-9}
		~& Correlation & slope & intersection & reduced $\chi^{2}$ & Correlation & slope & intersection & reduced $\chi^{2}$ \\ \midrule
		~& ~ & ~ & ~ & $\lambda5780$ DIB & ~ & ~ & ~ & ~ \\ \midrule
		\midrule
		\dt          &  0.96  & 0.26  & 1.62   &  5.1      & 0.97  & 0.28  & 1.91       & 8.9    \\
		\na D$_{2}$  &  --    & --    &  --    &    --     & 0.81  & 0.38  & $-$9.50    & 11.7   \\
		\na D$_{1}$  &  --    & --    &  --    &    --     & 0.84  & 0.43  & $-$2.03    & 15.4   \\
		\ca          &  0.96  & 1.13  & 4.02   &  15.1     & 0.94  & 0.92  & 5.12       & 22.35  \\
		\bottomrule
		~& ~ & ~ & ~ & $\lambda5797$ DIB & ~ & ~ & ~ & ~ \\ \midrule
		\na D$_{2}$  &   --   &  --   &   --   &   --     & 0.80   & 0.11   &  $-$1.05  & 5.8    \\
		\na D$_{1}$  &   --   &  --   &   --   &   --     & 0.84   & 0.25   &  $-$13.12 & 6.24   \\
		\ca          &  0.96  &  0.32 & 0.34   &  4.2     & 0.90   & 0.27   &  5.2      & 12.13  \\
		\bottomrule
	\end{tabular}
	\label{t1a}
\end{table} 

\end{document}